\pdfoutput=1

\documentclass[11pt,twoside,a4paper,cmspaper,final,collab]{cms-tdr}
\def\svnVersion{426725P}\def\cmsCernNoTag{CERN-EP-2016-306}\def\cmsCernDate{\today}\def\cmsMessage{Published in the European Physical Journal C as \href{http://dx.doi.org/10.1140/epjc/s10052-017-5182-1}{\doi{10.1140/epjc/s10052-017-5182-1.}}}

\begin{document}\cmsNoteHeader{SUS-16-003}

\hyphenation{had-ron-i-za-tion}
\hyphenation{cal-or-i-me-ter}
\hyphenation{de-vices}
\RCS$HeadURL: svn+ssh://svn.cern.ch/reps/tdr2/papers/SUS-16-003/trunk/SUS-16-003.tex $
\RCS$Id: SUS-16-003.tex 422176 2017-08-23 09:53:15Z lshchuts $

\newcommand{\mllumi}{2.3\fbinv}
\providecommand{\MT}{\ensuremath{M_\mathrm{T}}\xspace}
\newcommand{\Njets}{\ensuremath{N_\mathrm{j}}\xspace}
\newcommand{\Nbjets}{\ensuremath{N_{\PQb}}\xspace}
\newcommand{\WZ}{\ensuremath{\PW\cPZ}\xspace}
\newcommand{\ttZ}{\ensuremath{\ttbar\cPZ}\xspace}
\newcommand{\ttW}{\ensuremath{\ttbar\PW}\xspace}
\newcommand{\ttH}{\ensuremath{\ttbar\PH}\xspace}
\newcommand{\ptRatio}{\ensuremath{\pt^\text{ratio}}\xspace}
\newcommand{\ptRel}{\ensuremath{\pt^\text{rel}}\xspace}
\newcommand{\miniIso}{\ensuremath{I_\text{mini}}\xspace}

\providecommand{\NA}{\ensuremath{\text{---}}\xspace}
\ifthenelse{\boolean{cms@external}}{\providecommand{\cmsLeft}{top\xspace}}{\providecommand{\cmsLeft}{left\xspace}}
\ifthenelse{\boolean{cms@external}}{\providecommand{\cmsRight}{bottom\xspace}}{\providecommand{\cmsRight}{right\xspace}}

\cmsNoteHeader{SUS-16-003}
\title{Search for new phenomena with multiple charged leptons in proton-proton collisions at $\sqrt{s}= 13$\TeV}

\date{\today}

\abstract{
Results are reported from a search for physics beyond the standard model in final states with at least three charged leptons, in any combination of electrons or muons. The data sample corresponds to an integrated luminosity of 2.3\fbinv of proton-proton collisions at $\sqrt{s} =13$\TeV, recorded by the CMS experiment at the LHC in 2015. Two jets are required in each event, providing good sensitivity to strong production of gluinos and squarks. The search regions, sensitive to a range of different new physics scenarios, are defined using the number of jets tagged as originating from bottom quarks, the sum of the magnitudes of the transverse momenta of the jets, the imbalance in the overall transverse momentum in the event, and the invariant mass of opposite-sign, same-flavor lepton pairs. The event yields observed in data are consistent with the expected background contributions from standard model processes. These results are used to derive limits in terms of $R$-parity conserving simplified models of supersymmetry that describe strong production of gluinos and squarks. Model-independent limits are presented to facilitate the reinterpretation of the results in a broad range of scenarios for physics beyond the standard model.}

\hypersetup{%
pdfauthor={CMS Collaboration},%
pdftitle={Search for new phenomena with multiple charged leptons in proton-proton collisions at sqrt(s) = 13 TeV},%
pdfsubject={CMS},%
pdfkeywords={CMS, physics, supersymmetry}}

\maketitle
\section{Introduction}
\label{sec:Introduction}
{\tolerance=1600
Many types of beyond-the-standard-model (BSM) theories can produce multilepton events (three or more leptons) with a wide array of unique signatures~\cite{Eboli:1998yi,Craig:2012vj,Craig:2016ygr,Craig:2012pu,Chen:2013xpa}, including a number of supersymmetric (SUSY) models~\cite{Ramond:1971gb,Golfand:1971iw,Neveu:1971rx,Volkov:1972jx,Wess:1973kz,Wess:1974tw,Fayet:1974pd,Nilles:1983ge,Martin:1997ns,Baer:1995va}.
In these models, multilepton final states can arise from the decay of multiple vector bosons, e.g., in \ttbar production
with $\PQt \to \PQc\PH$ followed by $\PH \to \PW\PW^*$ or $\PH \to \cPZ\cPZ^*$, or in strong production of pairs of squarks or gluinos, which often initiate complex decay chains that can result in multiple $\PW$ \text{and/or} $\cPZ$ bosons.
The standard model (SM) processes that produce this final state are also characterized by multiple bosons and are well-understood both 
theoretically~\cite{Lazopoulos:2008de,Kardos:2011na,Campbell:2012dh,Campbell:2013yla,Kulesza:2015vda,Broggio:2015lya,Grazzini:2017ckn,Cascioli:2014yka,Caola:2015psa,Campbell:2016ivq,Binoth:2008kt,Nhung:2013jta,Yong-Bai:2015xna,Yong-Bai:2016sal,Hong:2016aek} 
and experimentally~\cite{Aaboud:2016xve,Aaboud:2016yus,Khachatryan:2016tgp,Aad:2015zqe,Khachatryan:2016txa}.
\par}

This paper describes a search for new physics in final states with three or more leptons, electrons or muons, produced at the CERN LHC, 
in proton-proton ($\Pp\Pp$) collisions at a center-of-mass energy of 13\TeV, with the CMS detector.
The data correspond to an integrated luminosity of $\mllumi$ collected in 2015. The expected irreducible backgrounds come from diboson production (\WZ and {\cPZ}{\cPZ}) or other SM processes, including {\ttbar}{W}, {\ttbar}{\cPZ}, and {\ttbar}{\PH}. These backgrounds are modeled using Monte Carlo (MC) simulations that have appropriate corrections applied to match the behavior of reconstructed objects in data. Reducible backgrounds are processes that produce one or more misidentified or nonprompt leptons, \ie those that arise from jets or meson decays, that pass all reconstruction, identification, and isolation criteria. Estimates of the probabilities of observing misidentified or nonprompt leptons based on control samples in data are used.

As an example of the type of BSM models for which this search has sensitivity, we interpret the results of this analysis in the context of SUSY models that feature strong production of pairs of squarks (\PSq) or gluinos (\PSg).
In addition to multiple leptons, these models predict that events can contain multiple jets, b-tagged jets, and missing transverse momentum.
Searches probing similar models have been carried out by the ATLAS and CMS Collaborations using pp collisions at 8\TeV~\cite{Aad:2014pda,Aad:2014lra,Aad:2015mia,Aad:2015zva,Chatrchyan:1696925,Khachatryan:2014doa,Khachatryan:1976453,Chatrchyan:2013xsw,Khachatryan:2016zcu}, 
and at 13\TeV~\cite{Aad:2016tuk,Aaboud:2016zdn,Aad:2016qqk,Aad:2016eki,Khachatryan:2016kdk,Khachatryan:2016xvy,Khachatryan:2016uwr,Khachatryan:2016kod}. 
Previous searches
exclude models with gluino mass less than approximately
1500\GeV, for a neutralino mass of 50\GeV,
and models with bottom squark mass less than 830 \GeV.

The result of the search, which is consistent with SM expectation, can also be used to constrain other BSM models not explicitly considered in this paper. To this end, we also provide upper limits on possible BSM contributions in the kinematic tail of the search variables in terms of the product of cross section, detector acceptance, and selection efficiency.

\section{The CMS detector}
\label{sec:detector}
The CMS detector features a superconducting solenoid of 6\unit{m} internal diameter that creates a magnetic field of 3.8\unit{T}. Inside the magnet volume are a silicon pixel and strip tracker, an electromagnetic calorimeter (ECAL) made of lead tungstate crystals, and a hadron calorimeter (HCAL) made of brass and scintillator, each composed of a barrel and two endcap sections. Forward calorimeters provide additional pseudorapidity ($\eta$) coverage for the HCAL. Muons are detected in gas-ionization chambers embedded in the steel flux-return yoke outside the solenoid. The first level of the CMS trigger system, composed of specialized hardware processors, uses information from the calorimeters and muon detectors to select the most interesting events in a fixed time interval of less than 4\mus. The high-level trigger (HLT) processor farm further decreases the event rate from approximately 100\unit{kHz} to less than 1\unit{kHz}, before data storage. A more detailed description of the CMS detector, together with a definition of the coordinate system used and the relevant kinematic variables, can be found in Ref.~\cite{Chatrchyan:2008zzk}.
\section{Event selection and Monte Carlo simulation}
\label{sec:selection}
Events used in this analysis are selected by the triggers that collect dilepton and multilepton events for later study, using variables constructed by the HLT. One set of triggers requires two leptons satisfying loose isolation criteria and transverse momentum $\pt  > 17\GeV$ for the leading lepton and $\pt > 12\,(8)\GeV$ for the subleading lepton in the case of electrons (muons). The second set of triggers places no requirements on the isolation, has a lower \pt threshold for the two leptons, $\pt > 8 \GeV$, and also requires that the scalar sum of jets with $\pt > 40$\GeV reconstructed in the HLT be greater than 300\GeV.

Electron candidates are reconstructed using tracking and electromagnetic calorimeter information by combining Gaussian sum filter tracks and ECAL energy deposits \cite{Khachatryan:2015hwa}. The electron identification is performed using a multivariate discriminant built with shower shape, track cluster matching, and track quality variables. The working point for the selection is chosen to maintain approximately 90\% efficiency for accepting electrons produced in the decays of W and Z bosons and also to efficiently reject candidates originating from jets. To reject electrons originating from photon conversions, electrons are required to have hits in all possible inner  layers of the tracker and to be incompatible with any secondary vertices containing only another electron. The selected electron candidates must have $\abs{\eta} < 2.5$.

Muon candidates are reconstructed in a global fit to the combined information from both the silicon tracker and the muon spectrometer~\cite{Chatrchyan:2012xi}. An identification is performed using the quality of the geometrical matching between measurements in the tracker and the muon system. To ensure the candidates are within the fiducial volume of the detector, we require that the candidate pseudorapidities satisfy $\abs{\eta} <  2.4$.

The reconstructed vertex with the largest value of summed physics-object $\pt^2$ is taken to be the primary $\Pp\Pp$ interaction vertex. 
The physics objects are the objects returned by a jet finding algorithm~\cite{Cacciari:2008gp,Cacciari:2011ma} applied to 
all charged tracks associated with the vertex, plus the corresponding associated missing transverse momentum.
Both electron and muon candidates are required to have a transverse (longitudinal) impact parameter of less than 0.5 (1.0) mm from the primary vertex. In addition, a requirement on the three-dimensional impact parameter significance is applied. This variable is the value of the impact parameter divided by its uncertainty and is required to be less than 4 for both electrons and muons. The rejection of nonprompt leptons is more efficient using the impact parameter significance than the value of impact parameter for similar prompt-lepton acceptance.

Lepton isolation is constructed using three different variables. The mini isolation, $\miniIso$, is the ratio of the amount of measured energy in a cone to the transverse momentum of the lepton. The radius is \pt-dependent: $R_{\text{iso}} = 10 \GeV/\text{min}(\text{max}(\pt(\ell), 50\GeV), 200\GeV)$, resulting in radii between 0.05 and 0.2. Requiring $\miniIso$ to be below a given threshold ensures that the lepton is locally isolated, even in Lorentz-boosted topologies.

{\tolerance=1200
The second variable is the ratio of the lepton \pt and the \pt of the jet matched to the lepton: $\ptRatio=\pt(\ell) / \pt(\text{jet})$. This jet must be separated by no more than 0.4 in $\Delta R$ from the lepton it is matched to, where $\Delta R = \sqrt{\smash[b]{\Delta\phi^2 + \Delta\eta^2}}$. In most cases, this is the jet containing the lepton. If no jet is found within $\Delta R < 0.4$, then $\ptRatio = 1$. The use of \ptRatio is a simple way to identify nonprompt low-\pt leptons originating from low-\pt b-quarks that decay with larger opening angles than the one used in the mini isolation.
\par}

The last variable is \ptRel, which is calculated by subtracting the lepton momentum from the momentum vector of the geometrically matched jet described above and then finding the component of the lepton momentum that is transverse to this new vector. If there is no matched jet, $\ptRel = 0$. This variable allows us to recover leptons from accidental overlap with jets in events where some of the final state particles are close together in Lorentz-boosted topologies.

Using the three variables above, a lepton is considered isolated if $\miniIso < I_1$ and that either $\ptRatio > I_2$ or $\ptRel > I_3$.
The $I_i$ values depend on the flavor of the lepton. The probability to misidentify a jet is higher for electrons, so tighter isolation values are used.
The logic behind this isolation is that a lepton should be locally isolated ($\miniIso$) and should carry the major part of the energy of the corresponding jet (\ptRatio) unless its overlap with the jet is accidental (\ptRel).
For electrons (muons), the tight selection requirements are $I_1 = 0.12\,(0.16)$, $I_2 = 0.76\,(0.69)$, and $I_3 = 7.2\,(6.0)$\GeV. The loose lepton isolation is relaxed to $\miniIso < 0.4$, and the other requirements are dropped. The loose leptons are used for background estimates. 
These selection requirements were optimized using MC simulations.

The offline selection requires at least three well-identified leptons in the event and any pair of opposite sign and same flavor (OSSF) leptons having an invariant mass greater than 12\GeV to reject low mass Drell--Yan and quarkonium processes. The leptons must pass offline \pt thresholds of 20, 15, and 10\GeV for the first, second, and third lepton, respectively, when \pt-ordered. For this offline selection, the trigger efficiency is above 99\%.

Jets are reconstructed from particle-flow candidates~\cite{Sirunyan:2017ulk} clustered using the anti-\kt algorithm~\cite{Cacciari:2008gp} 
with a distance parameter of 0.4 as implemented in the \FASTJET package~\cite{Cacciari:2011ma}. 
Only jets with $\pt>30\GeV$ and within the tracker acceptance $\abs{\eta}<2.4$ are considered. 
Additional criteria are applied to reject events containing noise and mismeasured 
jets~\cite{Chatrchyan:2011ds,Khachatryan:2016kdb,CMS-PAS-JME-16-003}. To avoid double counting, the closest matching jets to leptons 
are not considered if they are separated from the lepton by less than 0.4 in $\Delta R$. From those selected jets, the quantity \HT 
is defined by $ \HT = \sum_\text{jets} |\vec{\pt}|$, for all jets that satisfy the above-mentioned criteria. Jet energies are corrected for a shift in the energy scale, contributions from additional, simultaneous $\Pp\Pp$ collisions (pileup), and residual nonuniformity and nonlinearity differences between data and simulation~\cite{Khachatryan:2016kdb}.

The combined secondary vertex algorithm~\cite{CMS-BTV-12-001,CMS-PAS-BTV-15-001} is used to assess the likelihood that a jet originates from a bottom quark (``b jet''). Jets in this analysis are considered to be b tagged if they pass the algorithm's medium working point, which has a tagging efficiency of approximately 70\% and a mistag rate of approximately 1\% for light quarks and gluons.

{\tolerance=1200
The missing transverse momentum \ptvecmiss is defined as the negative vector sum of transverse momenta of all particle-flow candidates reconstructed in the event. Its magnitude is referred to as \ptmiss. Jet energy corrections are propagated to the \ptmiss following the procedure described in Ref.~\cite{Khachatryan:2014gga}.
\par}

To estimate the contribution of SM processes to the signal regions (described in Section~\ref{sec:strategy}) and to calculate the efficiency for new physics models, MC simulations are used. All the SM samples are generated using the \MGvATNLO2.2.2~\cite{MADGRAPH5,Alwall:2007fs,Frederix:2012ps} program at leading order (LO) or next-to-leading order (NLO) in perturbative QCD, with the exception of the diboson production samples (\WZ and {\cPZ}{\cPZ}) that are generated using \POWHEG v2~\cite{Nason:2004rx,Frixione:2007vw,Alioli:2010xd,Melia:2011tj,Nason:2013ydw} at NLO precision. The NNPDF3.0~\cite{Ball:2014uwa} LO (NLO) parton distribution function (PDF) set is used in MC simulations generated at LO (NLO). 
Parton showering and hadronization are simulated using \PYTHIA8.205~\cite{Sjostrand:2014zea} with the underlying event tune CUETP8M1~\cite{CMS-PAS-GEN-14-001}. 
The CMS detector response is determined using a \GEANTfour-based model~\cite{Geant}.

Events corresponding to the production of SUSY processes are generated with \MGvATNLO at LO precision, allowing up to two additional partons in the matrix element calculations. The SUSY particle decays, parton showering, and hadronization are simulated with \PYTHIA. The detector response for signal events is simulated using a CMS fast-simulation package~\cite{1742-6596-219-3-032053} that is validated against the \GEANTfour-based model.  Cross sections for SUSY signal processes, calculated at NLO with next-to-leading-log (NLL) gluon resummation, are taken from the LHC SUSY Cross Section Working Group~\cite{Kulesza:2008jb,Kulesza:2009kq,Beenakker:2009ha,Beenakker:2011fu,Kramer:2012bx,Borschensky:2014cia}. All simulated events are processed with the same reconstruction procedure as data. They include the effects of additional interactions, which can occur in the same or adjacent beam crossings (pileup). The distribution of additional interactions is matched to that observed in data. The pileup interactions are simulated by overlaying the primary interaction with additional minimum bias events, which are generated with the same \PYTHIA configuration as described above.

\section{Search strategy}
\label{sec:strategy}
\begin{figure}[tbp!]
\centering
 	\includegraphics[width=0.32\textwidth]{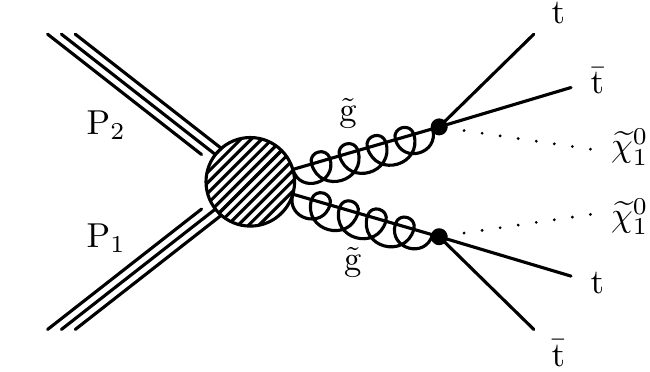}
	\includegraphics[width=0.32\textwidth]{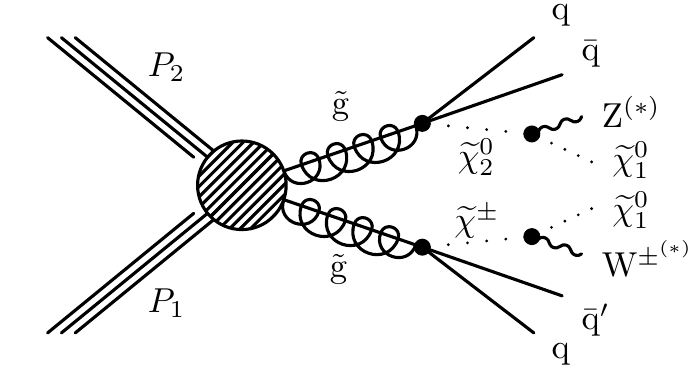}
	\includegraphics[width=0.33\textwidth]{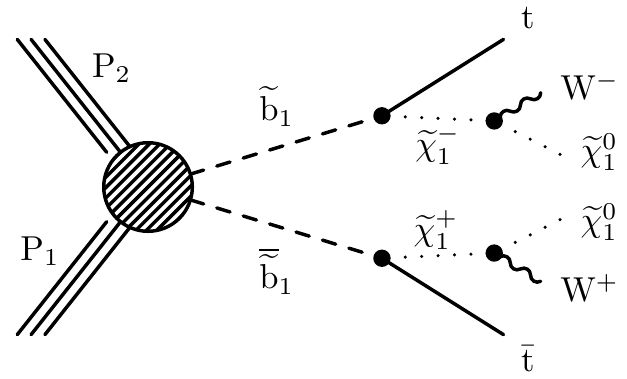}
\caption{\label{fig:diagrams}Diagrams for gluino and bottom squark pair production leading to multilepton events for simplified models of supersymmetry: (\cmsLeft) T1tttt, (middle) T5qqqqWZ, and (\cmsRight) T6ttWW.}
\end{figure}

The goal of this analysis is to search for possible excesses over the expected yields from SM processes in different categories of events with three or more leptons. With the \mllumi data sample at $\sqrt{s} = 13\TeV$, the search is focused on strongly produced SUSY particles, which benefit most from the increase of the production cross section with respect to 8 TeV. A few examples of diagrams of simplified models of SUSY processes~\cite{Alves:2011wf,Chatrchyan:2013sza} that can give rise to multilepton final states are shown in Fig.~\ref{fig:diagrams}. In these models, SUSY particles that are not directly included in the diagrams are assumed to be too heavy to be accessible at the LHC. Therefore, the free parameters in these models are usually the mass of the produced particles: gluinos and squarks, as well as the mass of the lightest supersymmetric particle (LSP).

Typical SUSY processes relevant for this work include T1tttt, which corresponds to gluino pair production where each gluino decays to a $\ttbar$ pair and 
the LSP (Fig.~\ref{fig:diagrams}-\cmsLeft). Another model, referred to as T5qqqqWZ, involves gluino pair production, where each gluino decays to a pair of light quarks (u, d, s, and c) and a neutralino ($\PSGcz_2$) or chargino ($\PSGcpm_1$), followed by decay of the neutralino or the chargino to a W or Z boson, respectively, and the LSP (Fig.~\ref{fig:diagrams}-middle). 
The probability for the decay to proceed via $\PSGcp_1$, $\PSGcm_1$, or $\PSGcz_2$ is 1/3 for each case, 
leading to the probabilities of having \PW\PW, \cPZ\cPZ\ or \PW\cPZ\ bosons in the final state to be about 44.5\%, 11.1\%, and 44.5\%, respectively. 
Only the final state with \PW\cPZ\ bosons contributes significantly to the acceptance of this search.  
Final states with \PW\PW\ bosons do not contribute, and the contribution from \cPZ\cPZ\ final states decaying to four leptons is negligibly small. 
In this scenario the neutralino and chargino are assumed to be mass-degenerate. A model called T6ttWW, features bottom squark pair production with their subsequent cascade decays via top quarks and W bosons (Fig.~\ref{fig:diagrams}-\cmsRight). The LSP is a neutralino in all of these models.

For the definition of the signal regions (SRs) we use several event variables: the number of b-tagged jets ($\Nbjets$), $\HT$, $\ptmiss$, and a classification depending on whether the event contains any OSSF dilepton pairs with an invariant mass between 76 and 106 \GeV, i.e. consistent with the Z boson (called ``on-Z'' if so and ``off-Z'' otherwise in the following). Events that do not contain any OSSF pairs are included in the off-Z sample.

The separation in b-tagged jet multiplicities maximizes signal-to-background ratios for different signal models. For example, the T1tttt model features several b jets, which would be categorized into SRs which are almost free of WZ background owing to the b-tagged jet requirement. Including the zero b-tagged SRs keeps the analysis sensitive to signatures such as the T5qqqqWZ model. Additionally, a categorization in $\HT$ and $\ptmiss$ is useful to distinguish between compressed and noncompressed SUSY spectra, i.e.\ models with small or large mass differences between the SUSY particles in the decay chain.

A baseline selection is applied to the data set to select events of interest: three or more electrons or muons satisfying the requirements $\pt \geq 20$, 15, and 10\GeV; $m_{\ell\ell} \geq 12\GeV$; at least two jets; $\HT \geq 60\GeV$; and $\ptmiss \geq 50\GeV$. Events containing additional leptons with $\pt > 10 \GeV$ are included in the event selection.
Table~\ref{table:srdefinition}  shows the definition of the subdivision of the baseline selection into two sets of SRs for events that contain on-Z and off-Z dilepton pairs.  There are 15 SRs for each of the two groups, hence in total 30 SRs.
A set of four SRs with low or medium $\HT$ and low or medium $\ptmiss$ are defined for each of the b-tagged jet multiplicities 0, 1, and 2. Motivated by the low expected yield of events with $\Nbjets \ge 3$, SR 13 is defined for high b-tagged jet multiplicities and also has $\ptmiss<300\GeV$ and $\HT<600\GeV$. Two additional SRs with large $\HT$ (SR 14) and large $\ptmiss$ (SR 15), respectively, have been defined as nearly background-free SRs, since noncompressed SUSY models can yield events with very large values of $\ptmiss$ or $\HT$. Both of these SRs are inclusive in the number of b-tagged jets, and every selected event with $\ptmiss \geq 300\GeV$ is categorized in SR 15, while SR 14 is populated with events with $\ptmiss < 300\GeV$ and $\HT \geq 600\GeV$.

\begin{table*}[tbp!]
\centering
\topcaption{Definition of multilepton signal regions. These regions are the same for the on-Z and off-Z regions.}
\label{table:srdefinition}
\begin{tabular}{c|ccc|c|c}
\hline
\multicolumn{1}{c}{\Njets} & \Nbjets                    & $\ptmiss$ (\GeVns{})             	    	 & $60 \leq \HT < 400\GeV$ 	      & \multicolumn{1}{c}{$400 \leq \HT < 600\GeV$}         & \multicolumn{1}{c}{$\HT \geq 600\GeV$ }   	\\ 	\hline
 \multirow{8}{*}{$\geq$2}     &  \multirow{2}{*}{0 } & 50--150  &    SR 1       &     SR 3        & \multirow{7}{*}{SR 14} 			\\	
  &                        & 150--300 &           SR 2       &     SR 4        &                                       					\\[1ex]
& \multirow{2}{*}{1}  & 50--150  &          SR 5       &     SR 7        &                                       					\\	
   &                       & 150--300 &         SR 6       &     SR 8        &                                       					\\[1ex]
& \multirow{2}{*}{2 } & 50--150  &       SR 9       &     SR 11       &                                        					\\	
    &                      & 150--300 &         SR 10       &    SR 12       &                                        					\\ \cline{4-5}
&$\geq$3  	      & 50--300  & 	   \multicolumn{2}{c|}{SR 13 }   &                               					\\ \cline{4-6}
& $\geq$0                 & $\geq$300       &  	   \multicolumn{3}{c}{SR 15 }                                             			\\
\hline
\end{tabular}
\end{table*}

\section{Background estimation}
\label{sec:backgrounds}
Backgrounds in the multilepton final states can be divided in three categories:
\begin{enumerate}
\item Nonprompt or misidentified leptons are those arising from heavy-flavor decays, misidentified hadrons,
  electrons from unidentified photon conversions, or muons from light-meson decays in flight.
For this analysis, $\ttbar$ events can enter the SRs if nonprompt leptons are present in addition to the prompt leptons from the W boson decays. 
These nonprompt leptons typically originate from semileptonic decays of hadrons containing a b quark, which, in this case, is not reconstructed as a jet. 
Therefore, $\ttbar$ events typically have low $\HT$ and $\ptmiss$ and predominately populate SR 1 and SR 5, with 0 and 1 b-tagged jets, respectively. 
  
  In addition to \ttbar, Drell--Yan events can enter the baseline selection, although they are largely suppressed by the $\ptmiss > 50$ \GeV requirement.
Processes that yield only one prompt lepton, e.g. W+jets and single top quark production, are effectively suppressed by the three-lepton requirement because of the low probability that the two nonprompt leptons pass the tight identification and isolation requirements.
\item Diboson production could yield multilepton final states with up to three prompt leptons in WZ production and up to four prompt leptons in ZZ production. Especially in signal regions without b-tagged jets, WZ production has a sizable contribution. The normalization of this background is obtained from a dedicated control region enriched in WZ events.
\item Other SM processes that can yield three or more leptons are \ttW, \ttZ, and triboson production VVV where V stands for a W or $\cPZ$ boson. We also include the contribution from the SM Higgs boson produced in association with a vector boson or a pair of top quarks in this category of backgrounds. Processes that produce additional leptons from internal conversions, which are events that contain a virtual photon that decays to leptons, are also included here as X+$\gamma$, where X is predominantly $\ttbar$ or $\cPZ$. Those backgrounds are obtained from simulation and appropriate systematic uncertainties are assigned.
\end{enumerate}

The background contribution from nonprompt and misidentified leptons is estimated using the ``tight-to-loose ratio'' method~\cite{Khachatryan:2016kod}. The tight-to-loose ratio $f$ is the probability for a nonprompt lepton that satisfies the loose requirements to also satisfy the full set of requirements. The nonprompt background contribution is obtained from the number of events in an application region containing events with at least one of the leptons failing the full set of tight identification and isolation requirements, but passing the loose requirements, weighted by $f / (1-f)$. This ratio is measured in a control sample of QCD multijet events that is enriched in nonprompt leptons (measurement region), by requiring exactly one lepton passing the loose object selection and one recoiling jet with $\Delta R(\text{jet},\ell)>1.0$.
To suppress events with leptons from W and Z boson decays, $\ptmiss < 20 \GeV$ and $\MT <20 \GeV$ are also required, where $\MT = \sqrt{\smash[b]{2\ptmiss \pt(\ell) (1-\cos{\Delta\phi})}}$ and $\Delta\phi$ is the difference in azimuthal angle between the lepton and \ptvecmiss. The remaining contribution from these electroweak processes within the measurement region is subtracted using estimates from MC simulations.

The dependence of the tight-to-loose ratio on the flavor of the jet from which the nonprompt lepton originates is reduced by parameterizing the ratio as a function of a variable that is more strongly correlated with the parent parton \pt than with lepton \pt. This variable is calculated by correcting the lepton \pt as a function of the energy in the isolation cone around it. This definition leaves the \pt of the leptons passing the isolation requirement unchanged and modifies the \pt of those failing the requirement, so that it is a better proxy for the parent parton \pt and results in a flatter tight-to-loose ratio as a function of the parent parton \pt. The cone correction significantly improves the results of the method when applying it in simulation. The flavor dependence, which is much more important for the case of electrons, is also reduced by adjusting the loose object selection to obtain similar ratios for nonprompt electrons that originate from both light- and heavy-flavor jets. To avoid experimental biases, the tight-to-loose ratio is also measured as a function of $\eta$.

The tight-to-loose ratio method of estimating the nonprompt background is validated in a control region exclusive to our baseline selection with minimal signal contamination. This region is defined by having three tight leptons, one or two jets, $20 < \ptmiss < 50$\GeV, and an off-Z dilepton pair.
We find agreement of the order of 20\% between the predicted and observed yields in this control region in data, which validates the predictions and uncertainties of this method.

The WZ process is one of the main backgrounds in the regions with zero b-tagged jets. The relative contribution of this process in various SRs 
is estimated from the MC simulation at NLO, but the normalization is taken from a control region that is highly enriched for this process: three leptons pass nominal identification and isolation requirements, two leptons form an OSSF pair with $\abs{m_{\ell\ell} - m_\cPZ } < 15$\GeV, the number of jets is zero or one, the number of b-tagged jets is zero, $30 < \ptmiss < 100 \GeV$, and the $\MT$ of the third lepton (not in the pair forming the \cPZ) is required to be at least 50\GeV to suppress contamination from Drell--Yan processes. The expected WZ purity in the selected sample is 84\%.
Using this control region, we find that the WZ background predictions from simulation are consistent with data. The ratio between the prediction and data obtained with \mllumi 
of data is $1.13 \pm 0.17$. 
The uncertainty on the normalization of the WZ background includes the statistical uncertainty related to the event yield in the CR and a systematic component related to a small contamination of the CR due to other processes.

\section{Systematic uncertainties}
\label{sec:systematics}
Systematic uncertainties are characterized as either experimental, theoretical, or arising from the limited size of simulated event samples. These sources of uncertainties and their magnitudes are described below, and are summarized in Table~\ref{tab:systSummary}. The table also provides the effect of varying the uncertainties by $\pm1$ standard deviation (s.d.) on 
the signal and background yields. The jet energy scale uncertainty and the uncertainty in the b tagging efficiency are the only ones that can cause simulated events to migrate between signal regions.

The major experimental source of uncertainty is the knowledge of the jet energy scale (JES), which accounts for differences between kinematical variables from data and simulation and affects signal and background events that are taken from simulation samples \cite{Khachatryan:2016kdb,CMS-PAS-JME-16-003}. 
For the data set used in this analysis, the uncertainties on the JES vary from 2 to 8\%, depending on the \pt and $\eta$ of the jet. The impact of these uncertainties is assessed by shifting the jet
energy correction factors for each jet up and down by $\pm$1 s.d. and recalculating all of the kinematic quantities. The JES uncertainties are propagated to the missing transverse momentum and all variables derived from jets (numbers of jets and b-tagged jets, and \HT) used in this analysis; this propagation results in 1--20\% variation in the MC background estimation in the regions with higher data yields.

A similar approach is used for the uncertainties associated with the corrections for the $\PQb$ tagging efficiencies for light-, charm-, and bottom-flavour jets, which are parametrized as a function of $\pt$ and $\eta$~\cite{CMS-BTV-12-001,CMS-PAS-BTV-15-001}.
The variation of the scale factor correcting for the differences between data and simulation is at maximum 5--10\%, and leads to an effect of 1--20\% on the yields, depending on the SR and on the topology of the events under study.
If  one considers only highly populated SRs to get an overview of the main effects on the background yields, the bulk of the \ttW yield varies by $\sim$10\% and the \WZ yield by $\sim$13\%.

Lepton identification scale factors have been measured by comparing efficiencies in data and simulation using the ``tag-and-probe'' method~\cite{Khachatryan:2015hwa,Chatrchyan:2012xi} and are applied as a function of lepton $\pt$ and $\eta$. The corresponding uncertainties on the scale factors have been evaluated and are approximately 2\% for both electrons and muons. Trigger efficiency scale factors have been found to be very close to unity. An uncertainty of 3\% in the scale factors has, however, been assigned to cover the difference between trigger efficiencies measured in simulation over a large number of samples.

All these uncertainties related to corrections of the simulation (JES
corrections, b tagging efficiency scale factors, lepton identification
and trigger scale factors) have been estimated also for the fast simulation used for the signal samples. We propagate them to the expected signal yields
following the same procedure.

The uncertainties in the renormalization ($\mu_{\textrm R}$) and factorization scales ($\mu_\mathrm{F}$)
and the PDF are considered for some of the rare processes, namely \ttW, \ttZ, and
\ttH. Both the changes in the acceptance and cross sections due to those effects are taken into
account.

For the study of the renormalization and factorization scale uncertainties, variations up and down by a factor of two
with respect to the nominal values of $\mu_{\textrm R}$ and $\mu_\mathrm{F}$ are
considered.
The maximum difference in the yields with respect to the nominal case is observed when both scales
are varied simultaneously up and down. The effect on the overall cross section is
found to be about $13\%$ for \ttW and about $11\%$ for \ttZ. An additional
uncertainty in the acceptance corresponding to different
signal regions is included. This is found to be between 3 and 18\% depending on the SR and process.

The uncertainty related to the PDFs is estimated from the 100 NNPDF
3.0 replicas by computing the deviation with respect to the nominal
yields for each of them, and for each signal region (the cross section and
acceptance effects are considered together)~\cite{Butterworth:2015oua}. The root mean square of
the variations is taken as the value of the systematic
uncertainty. Since no significant variations among the different
signal regions are seen, a flat uncertainty of $3 (2) \%$  is applied
to the \ttW (\ttZ) background. This value also includes the deviation resulting from varying the strong coupling strength
$\alpha_{S}(M_{\cPZ})$, which is added in quadrature, and whose magnitude
is similar to or smaller than that of the PDF set uncertainty. For the \ttH process, the same uncertainties as estimated for \ttZ are
applied. A theoretical uncertainty of 50\% is assigned to the
remaining rare processes.

In signal samples, the uncertainty due to initial-state radiation is computed as a function of the $\pt$ of the gluino pair using the methods described in Ref.~\cite{Chatrchyan:2013xna}. 
For values below 400\GeV, no uncertainty is applied. For values between 400 and 600\GeV, a 15\% uncertainty is assigned, and above 600\GeV this uncertainty is increased to 30\%.

The limited size of the  generated MC samples represents an additional source of uncertainty.
The uncertainty in signal processes and backgrounds such as \ttW, \ttZ, and \ttH, is calculated from the number of MC events entering each of the signal regions.

{\tolerance=1200
For the nonprompt and misidentified lepton background, we assign several systematic uncertainties. The statistical uncertainty resulting from the limited number of events in the application region used to estimate this background contribution varies from 1 to 100\%. The regions where these uncertainties are large are generally regions where the overall contribution of this background is small. When no events are observed in the application region, the upper limit of the background expectation is set to 0.35, which is found by applying the most probable tight-to-loose ratio as if the application region contained an event count equal to the variance of a Poisson distribution with a mean of zero.
\par}

The systematic uncertainties related to the extrapolation from the control regions to the SRs for the nonprompt lepton background are estimated to be 30\%. This magnitude has been extracted from the level of closure achieved in a test that was performed with MC samples yielding nonprompt leptons to validate background predictions based on control samples in data, as described in Section~\ref{sec:backgrounds}.

The uncertainty associated with the electroweak (EW) background subtraction in the tight-to-loose ratio computation is propagated through the full analysis process by replacing the nominal tight-to-loose ratio with another value obtained when the scale factor applied to the electroweak processes in the measurement region  is varied by 100\% of its difference from unity. 
The overall effect on the nonprompt background yield lies between 1 and 5\% depending on the SR considered.

The estimate of the \WZ\ background is assigned a 15\% normalization uncertainty using the measurement in a dedicated control region. This uncertainty is compatible with the one quoted for the experimental measurement of this process in Ref.~\cite{Khachatryan:2016tgp}. 
Additional uncertainties for the extrapolation from the control region to the signal regions of 10 -- 30\% are taken into account depending on the SR. These uncertainties are dominated by the JES and b tagging uncertainties described earlier.

Finally the uncertainty on the integrated luminosity is 2.7\% \cite{CMS-PAS-LUM-15-001}.

\begin{table*}[tbp!]
\centering
\topcaption{ Summary of the sources of uncertainties and their magnitudes. The third column provides the changes in yields of signal and background induced by one s.d. changes in the magnitude of uncertainties.\label{tab:systSummary}}
\resizebox{\textwidth}{!}{
\begin{tabular}{lcc}
\hline
Source          & Magnitude  (\%)   & Effect on yield (\%)\\
\hline
Integrated luminosity~\cite{CMS-PAS-LUM-15-001}      & $2.7$      & $2.7$ $^*$     \\[2ex]
Limited MC sample sizes & 1--100    & 1--100  $^*$     \\[2ex]
Jet energy scale          & 2--8    & 1--20 $^*$       \\
b tagging efficiency & 5--10   & 1--20  $^*$             \\
Pileup          & 5         & 3  $^*$                     \\[2ex]
Renormalization and factorization scales      & $-50$ / $+100$ &  11--13 (cross-section) / 3--18 (acceptance) (\ttW,\ttZ,\ttH)   \\
PDF            &  \NA           &  2--3 (\ttW,\ttZ,\ttH)      \\
Other backgrounds    & 50    & 50 (rare processes, tribosons, etc.)  \\[2ex]
Lepton efficiencies & $2$     & $6$ $^*$  \\
Trigger efficiencies & $3$      & $3$ $^*$ \\
FastSim lepton efficiencies & 3--10      & 3--10 FastSim signals \\
FastSim trigger efficiencies& $5$      & $5$ FastSim signals \\[2ex]
Tight-to-loose ratio control region statistical uncertainty     & 1--100      & 1--100 (nonprompt bkg. only)      \\
Tight-to-loose ratio systematic uncertainty& 30          & 30 (nonprompt bkg. only)          \\
EW subtraction in  tight-to-loose ratio & 100 (ewk. SF)  & 1--5 (nonprompt bkg. only)            \\
\WZ control region normalization     & 15        & 15 (\WZ only)        \\
\WZ extrapolation     & 10--30       & 2--30  (\WZ only)     \\
\hline
\multicolumn{3}{l}{$^*$ Applied to both signal and background simulation samples.} \\
    \end{tabular}
}
\end{table*}

\section{Results and interpretations}
\label{sec:results}
\begin{figure*}[!btp]
	\centering
    \includegraphics[width=.33\textwidth]{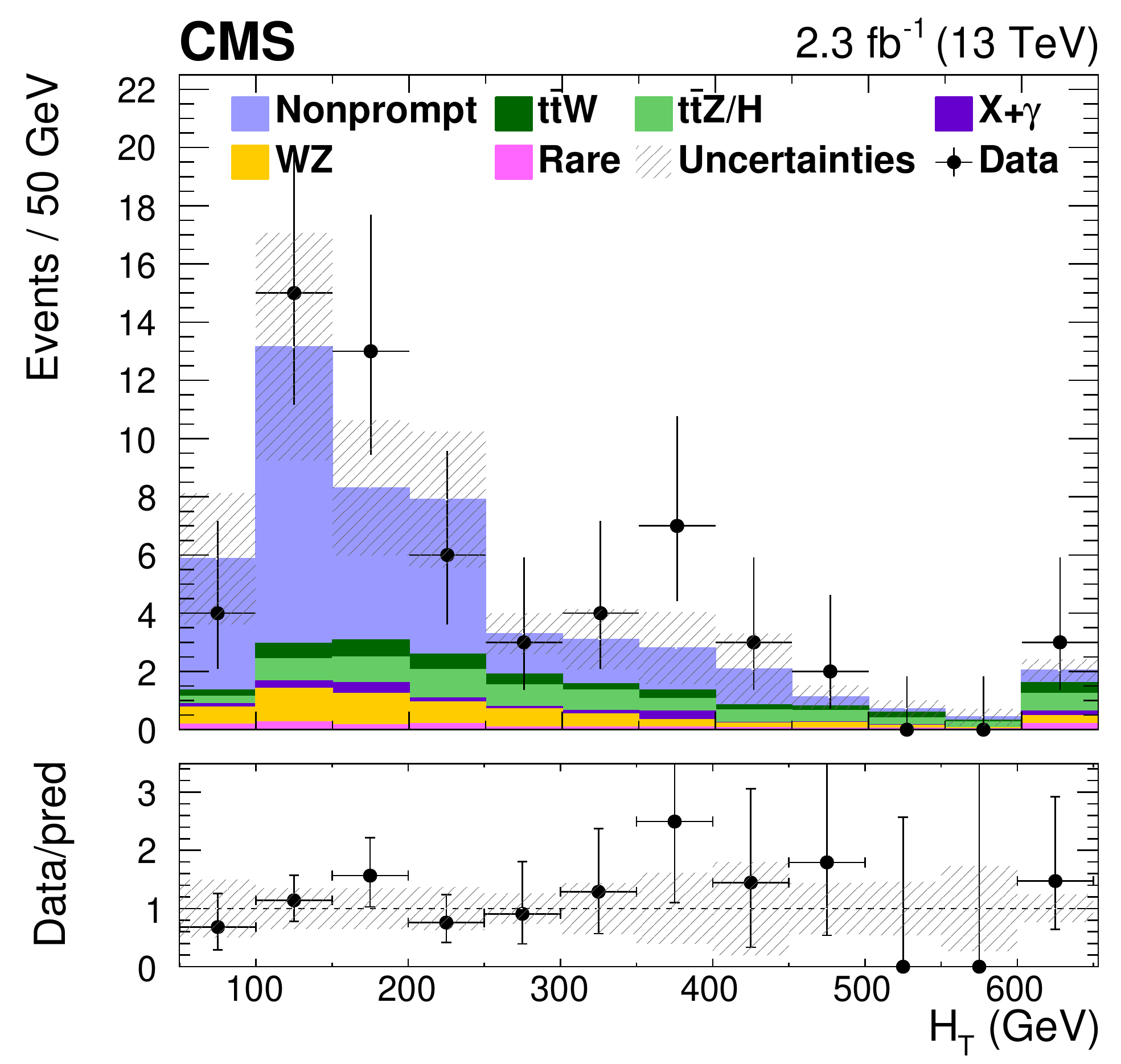}
	\includegraphics[width=.33\textwidth]{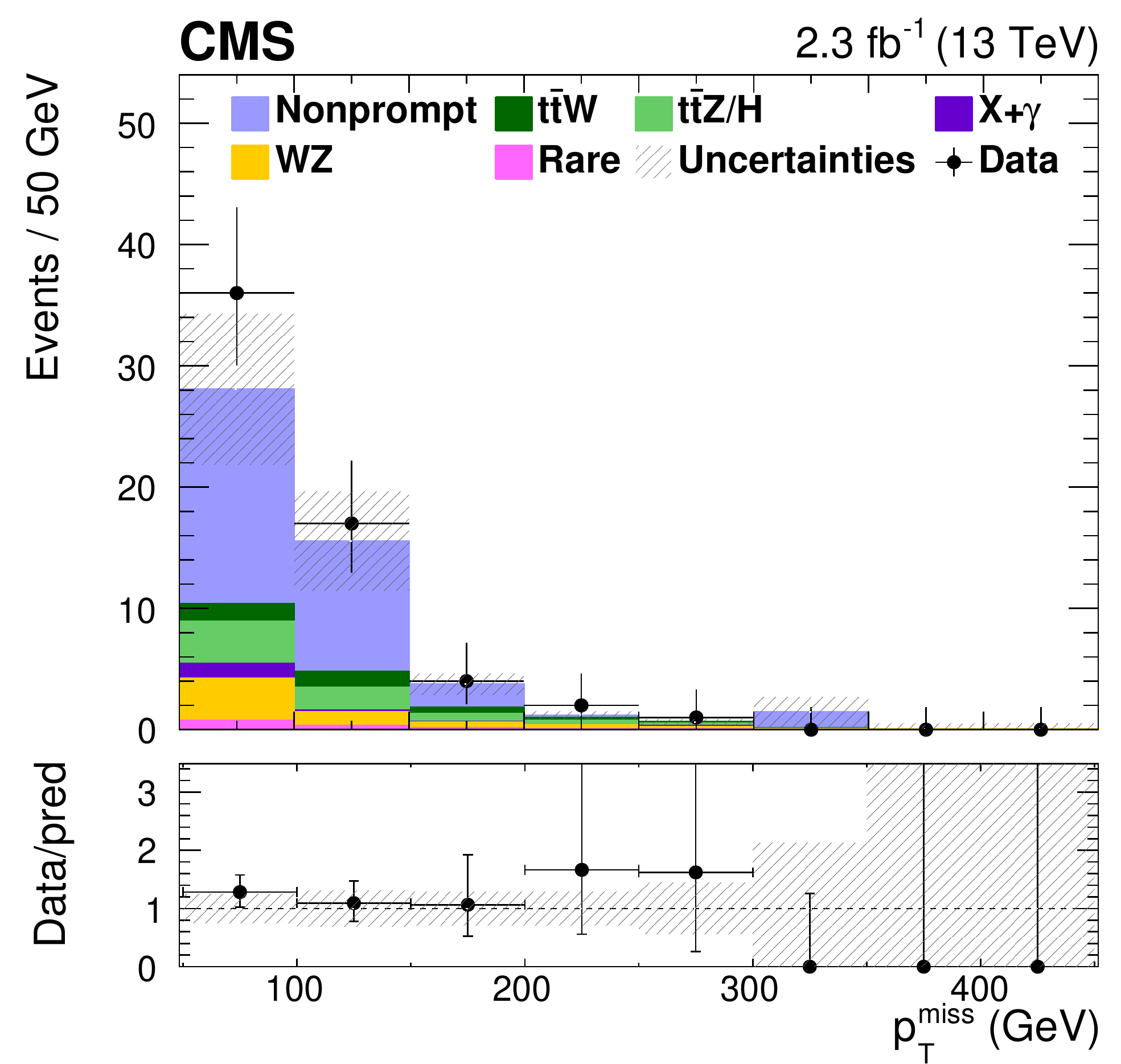}\\
	\includegraphics[width=.33\textwidth]{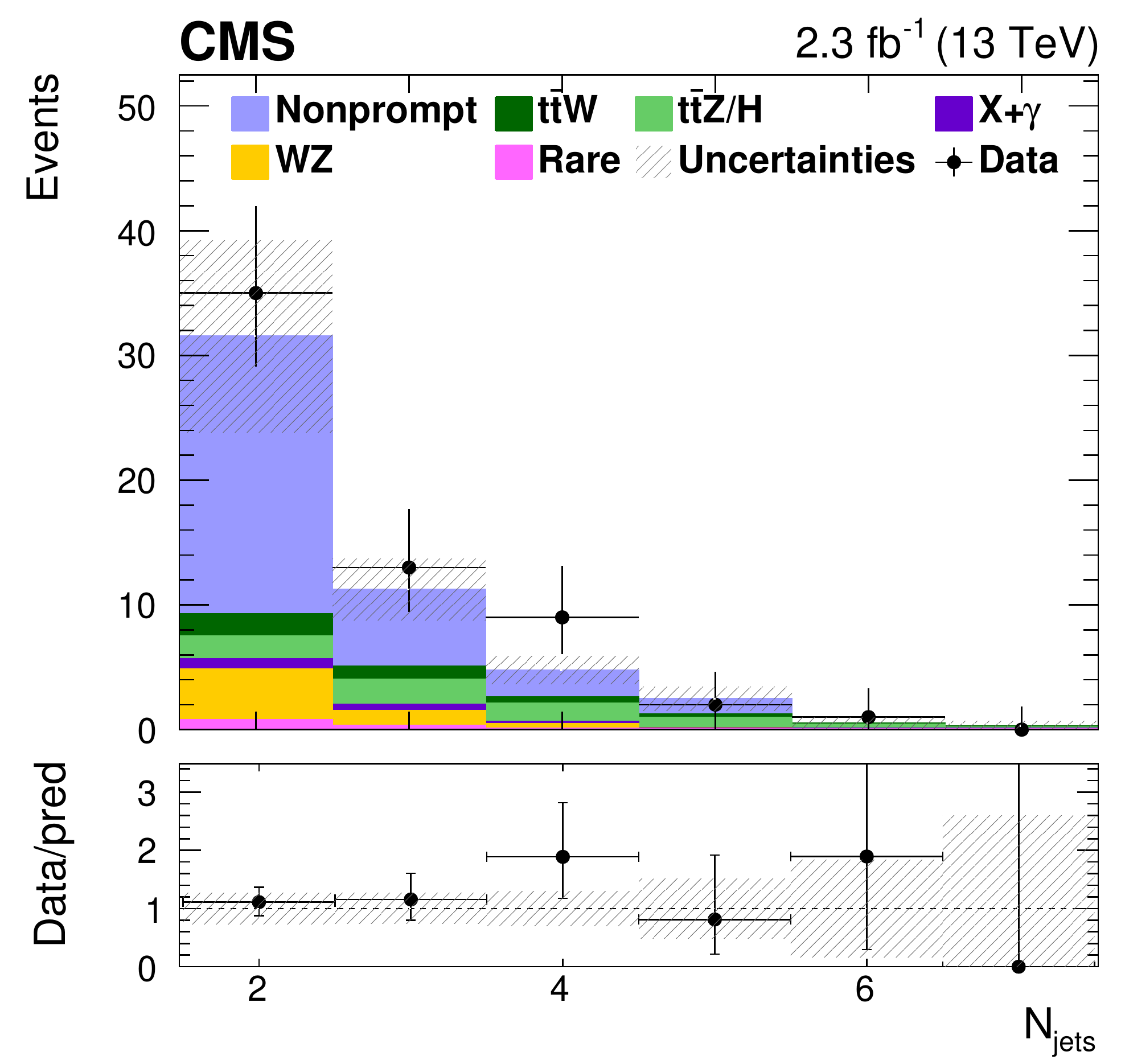}
	\includegraphics[width=.33\textwidth]{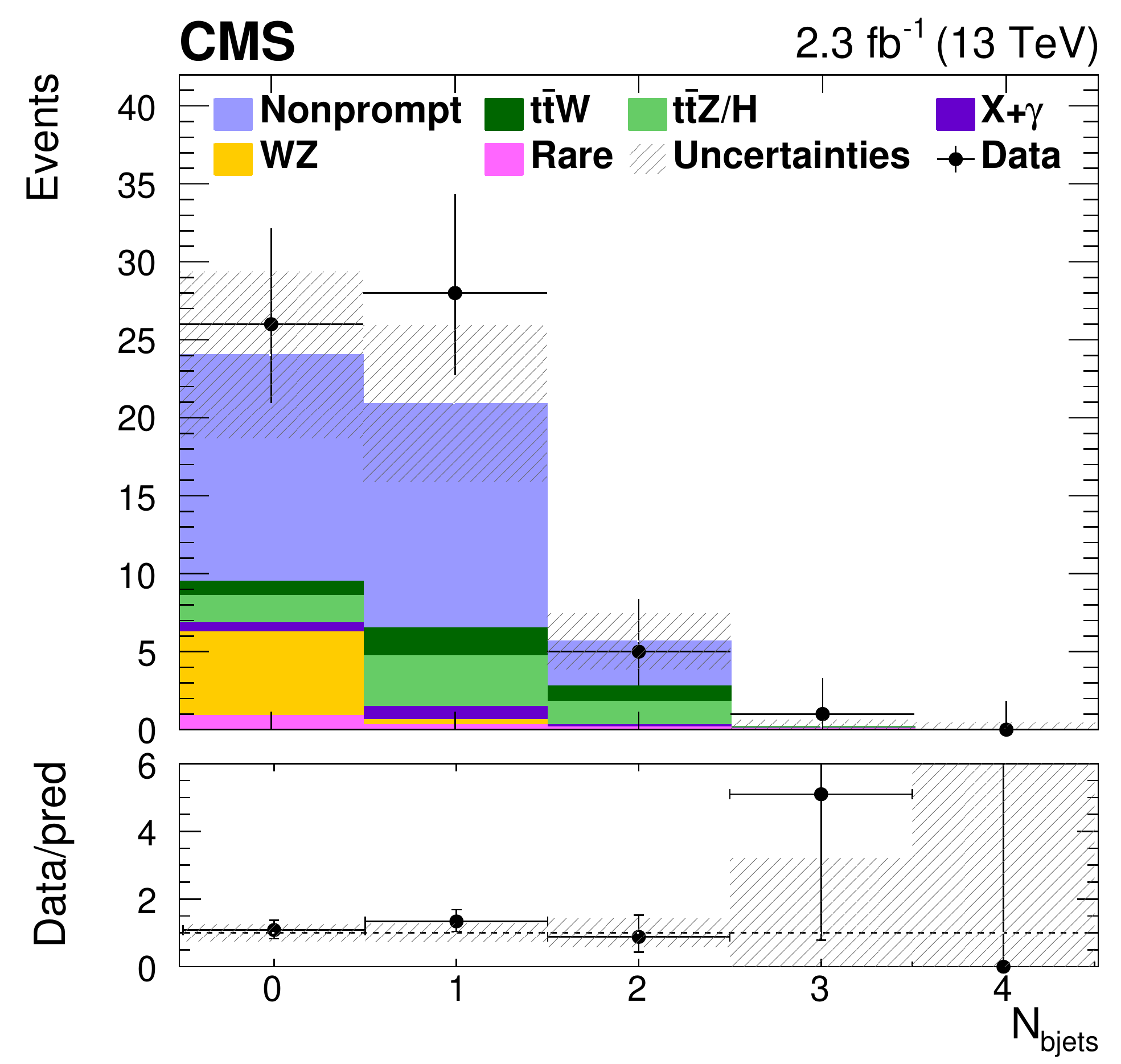}\\
	\includegraphics[width=.33\textwidth]{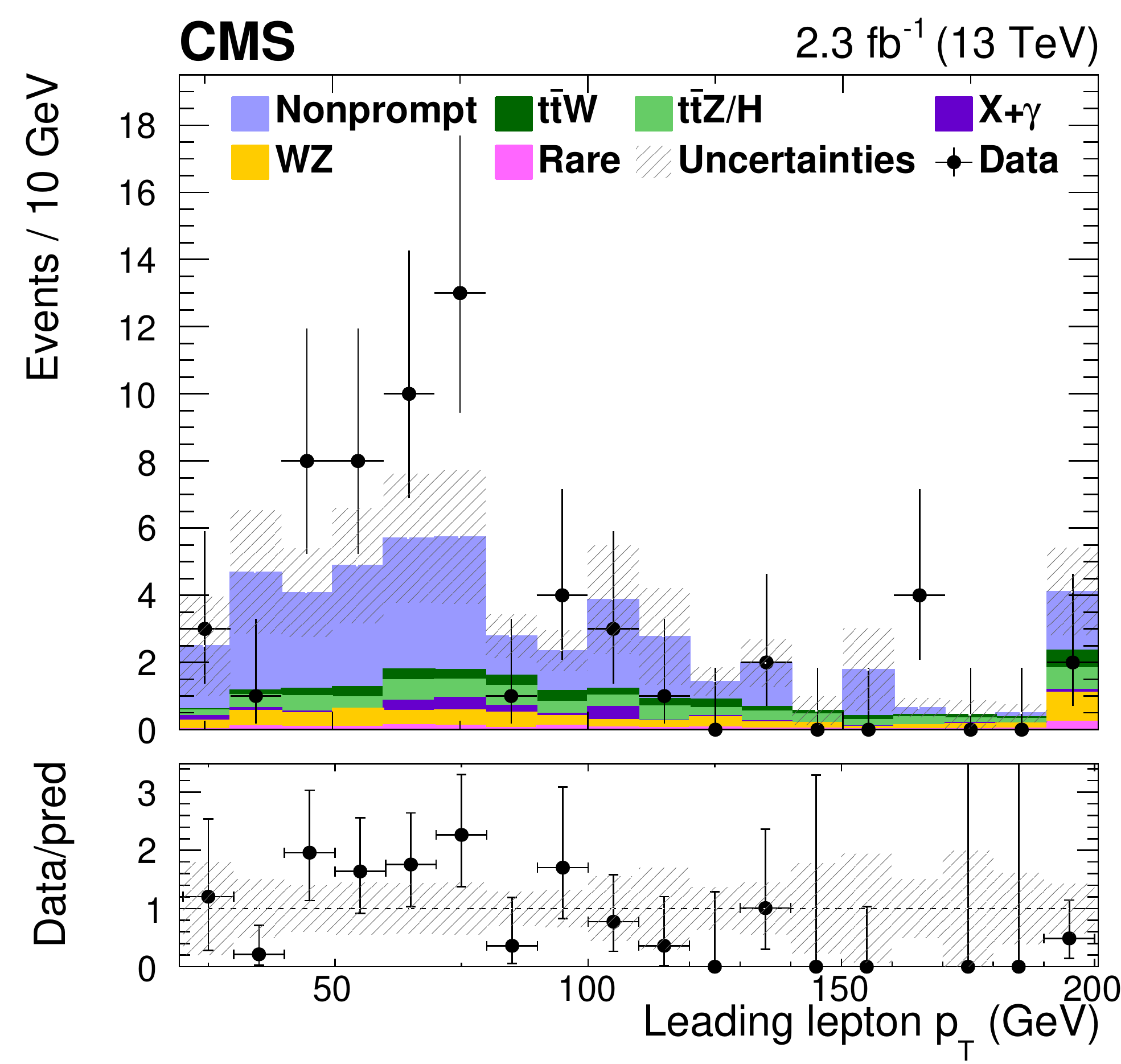}
	\includegraphics[width=.33\textwidth]{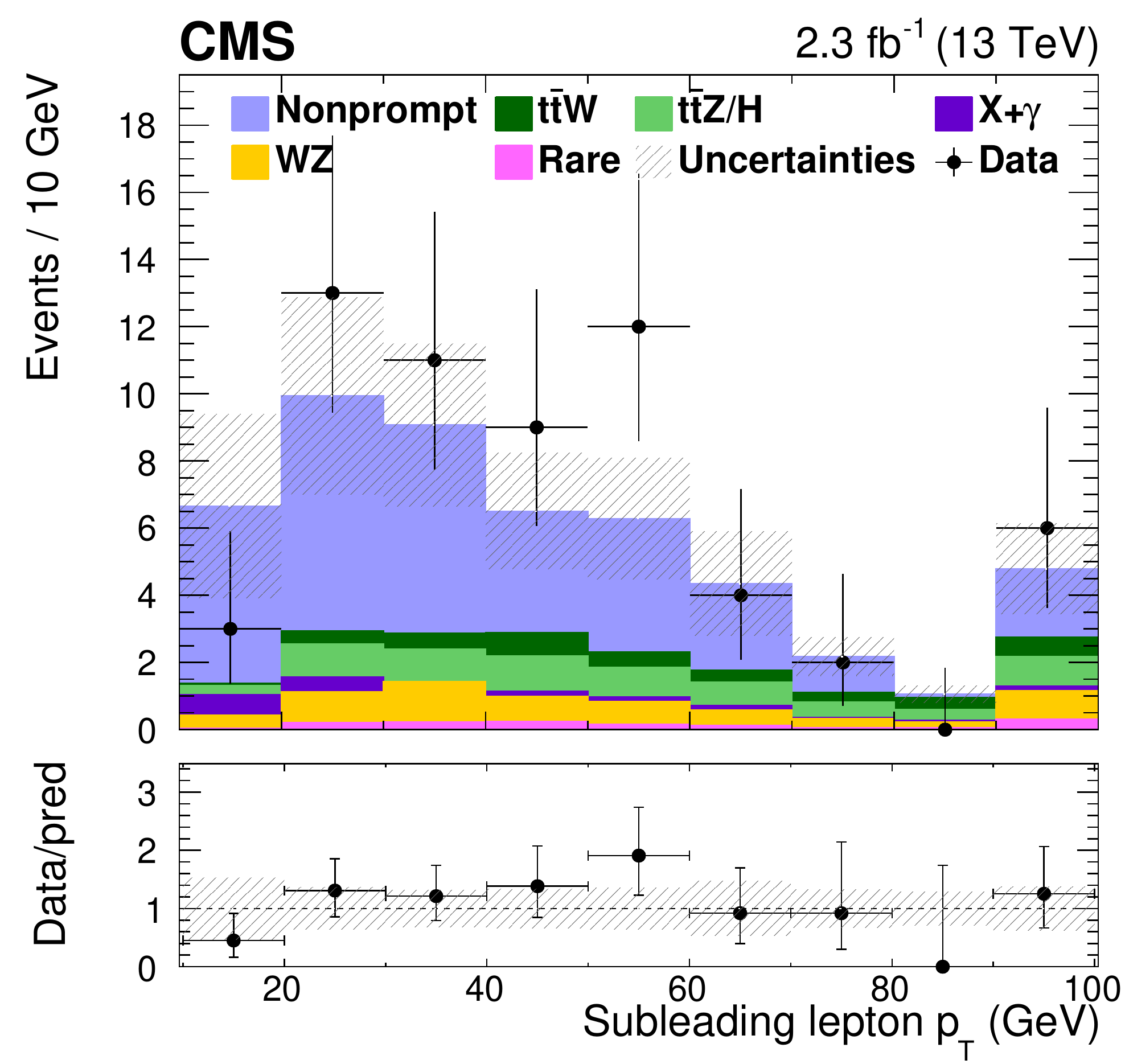}\\
	\includegraphics[width=.33\textwidth]{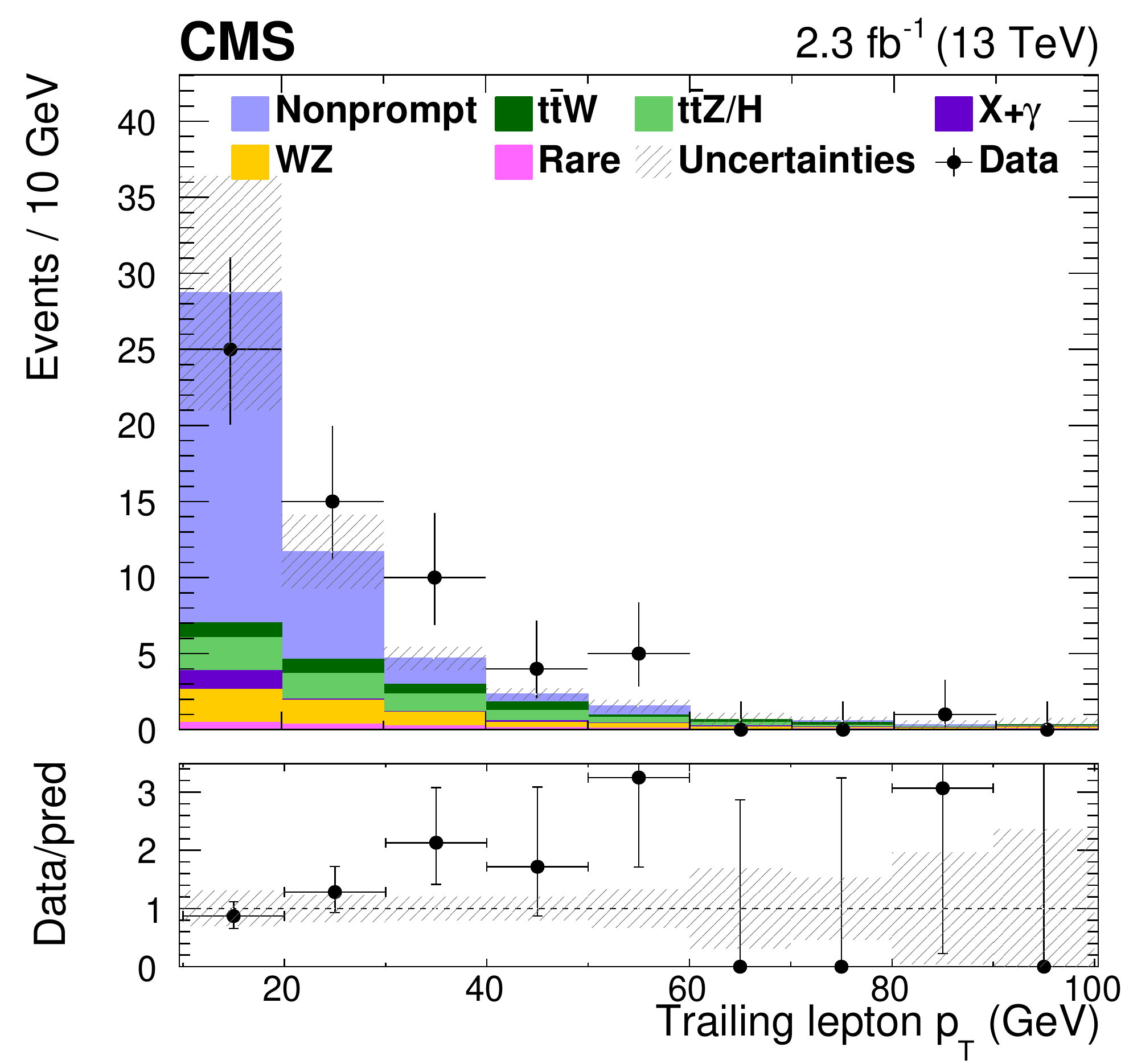}
	\includegraphics[width=.33\textwidth]{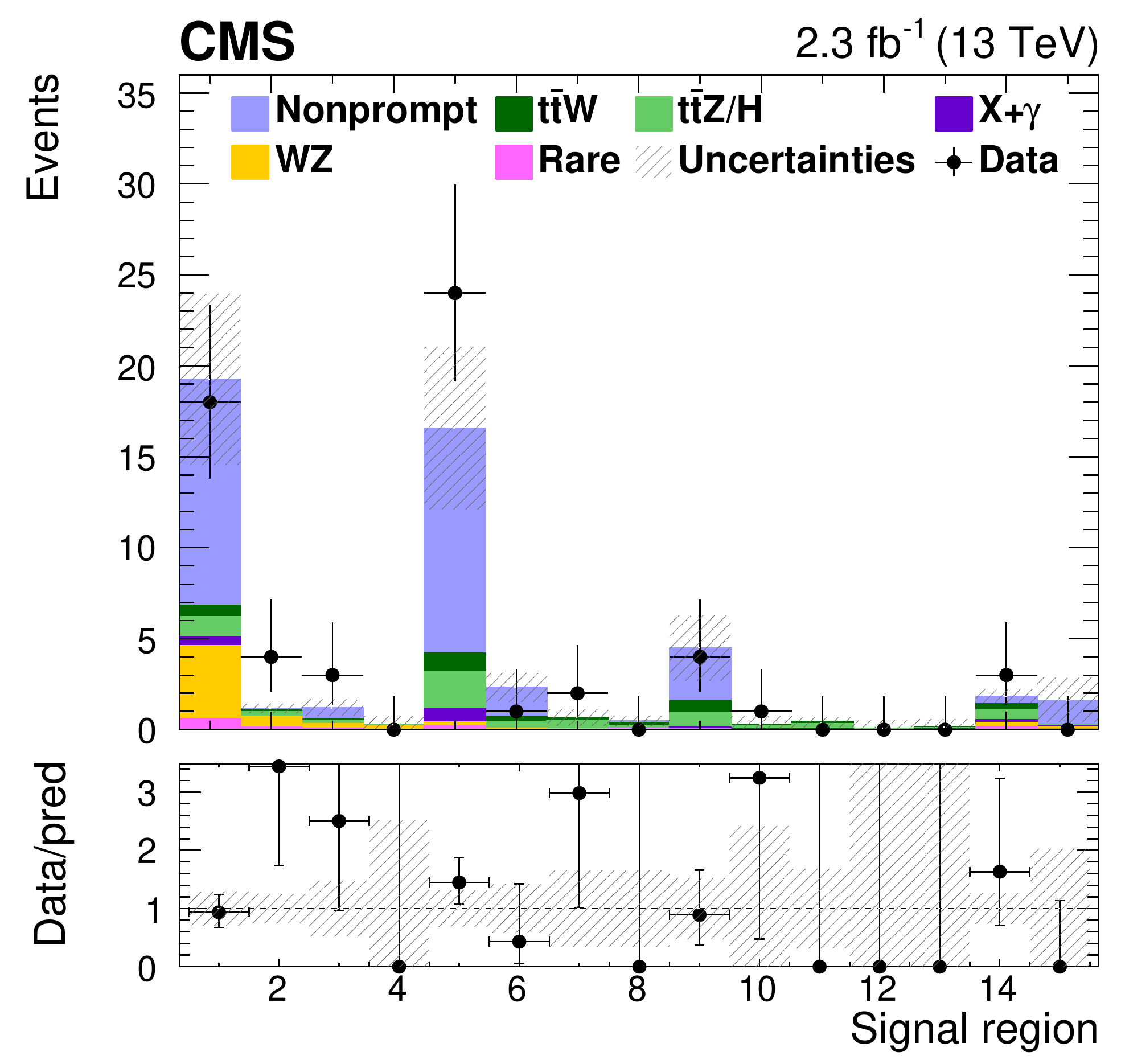}
\caption{Off-Z samples: from left to right, top to bottom, distributions of  $\HT$, $\ptmiss$,  $\Njets$, $\Nbjets$,  $\pt$ of leptons for the predicted backgrounds and for the data in the off-Z baseline selection region, in these plots the rightmost bin contains the overflow from counts outside the range of the plot. On the bottom-right corner the total predicted background and the number of events observed in the 15 off-Z SRs is shown.}
\label{fig:fullOffZ}
\end{figure*}

\begin{figure*}[!btp]
\centering
	\includegraphics[width=.33\textwidth]{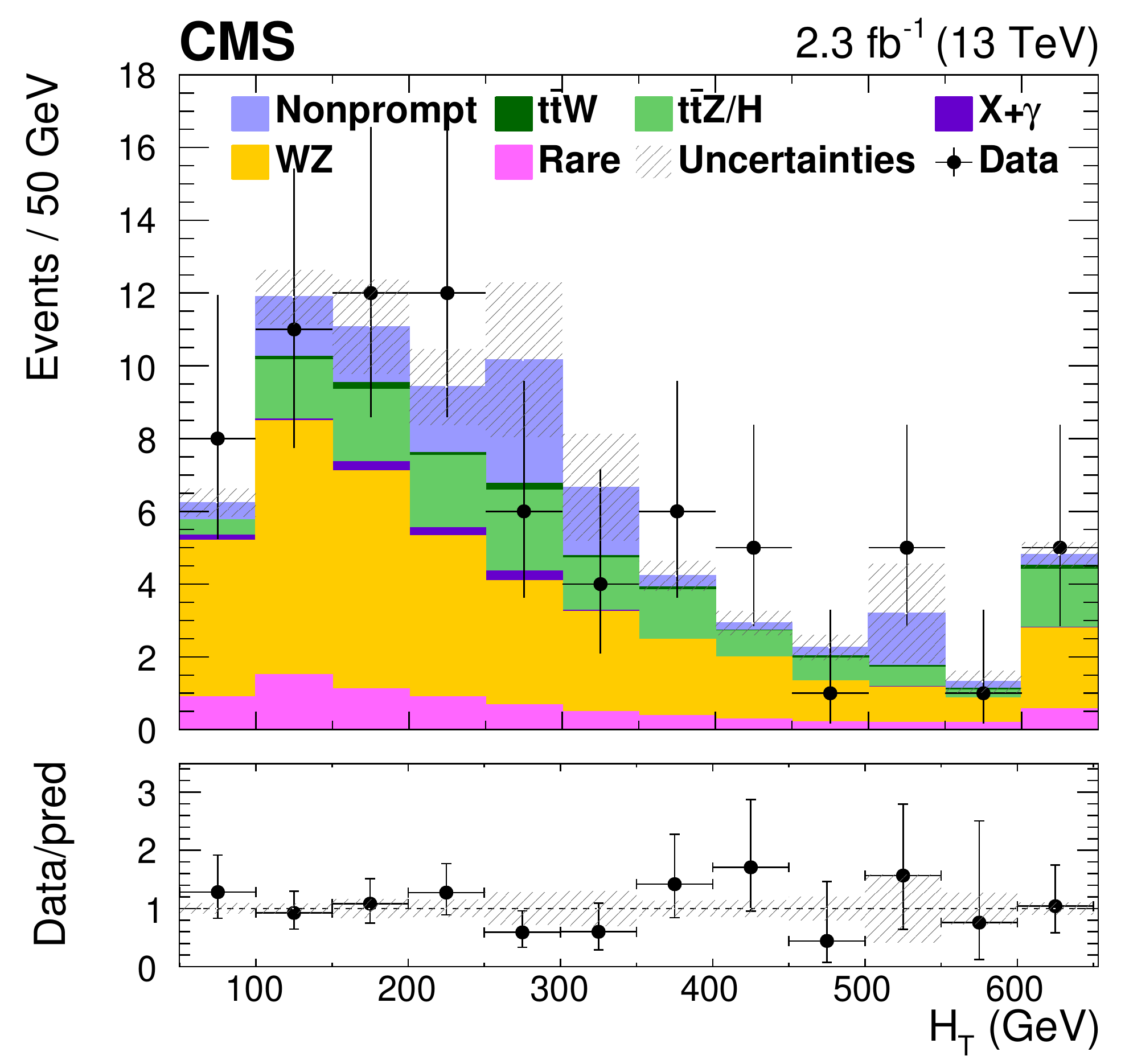}
	\includegraphics[width=.33\textwidth]{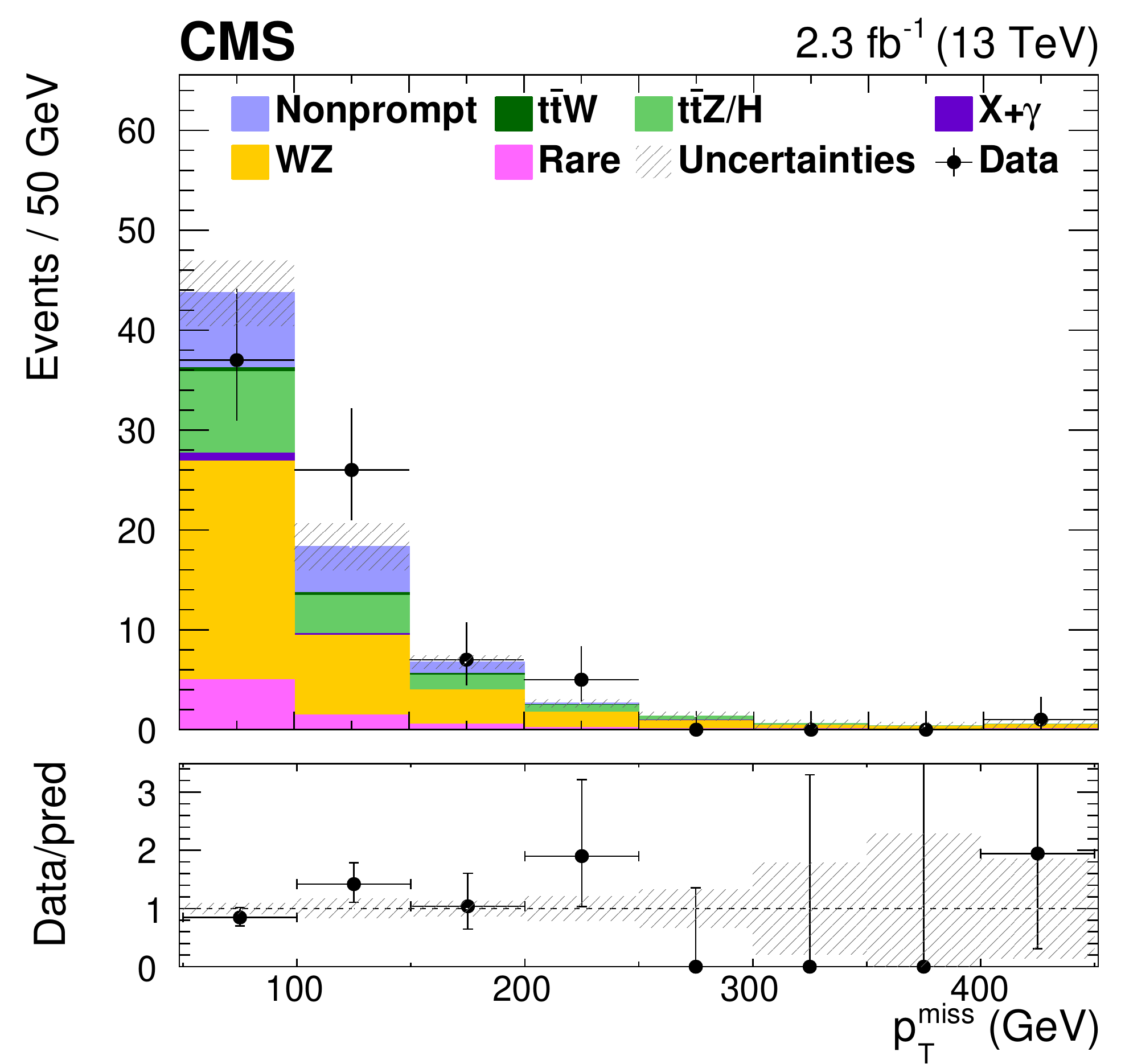} \\
	\includegraphics[width=.33\textwidth]{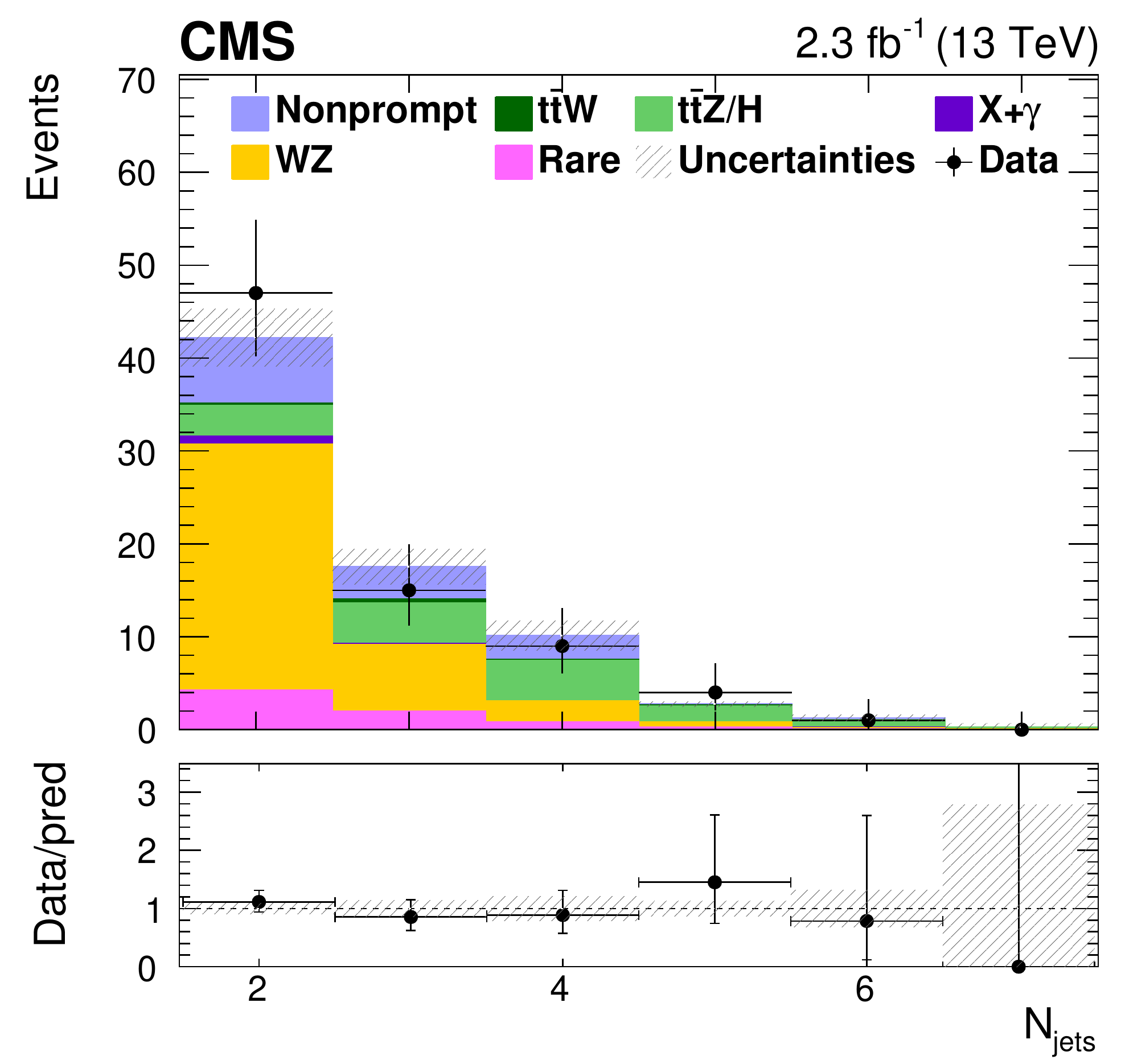}
	\includegraphics[width=.33\textwidth]{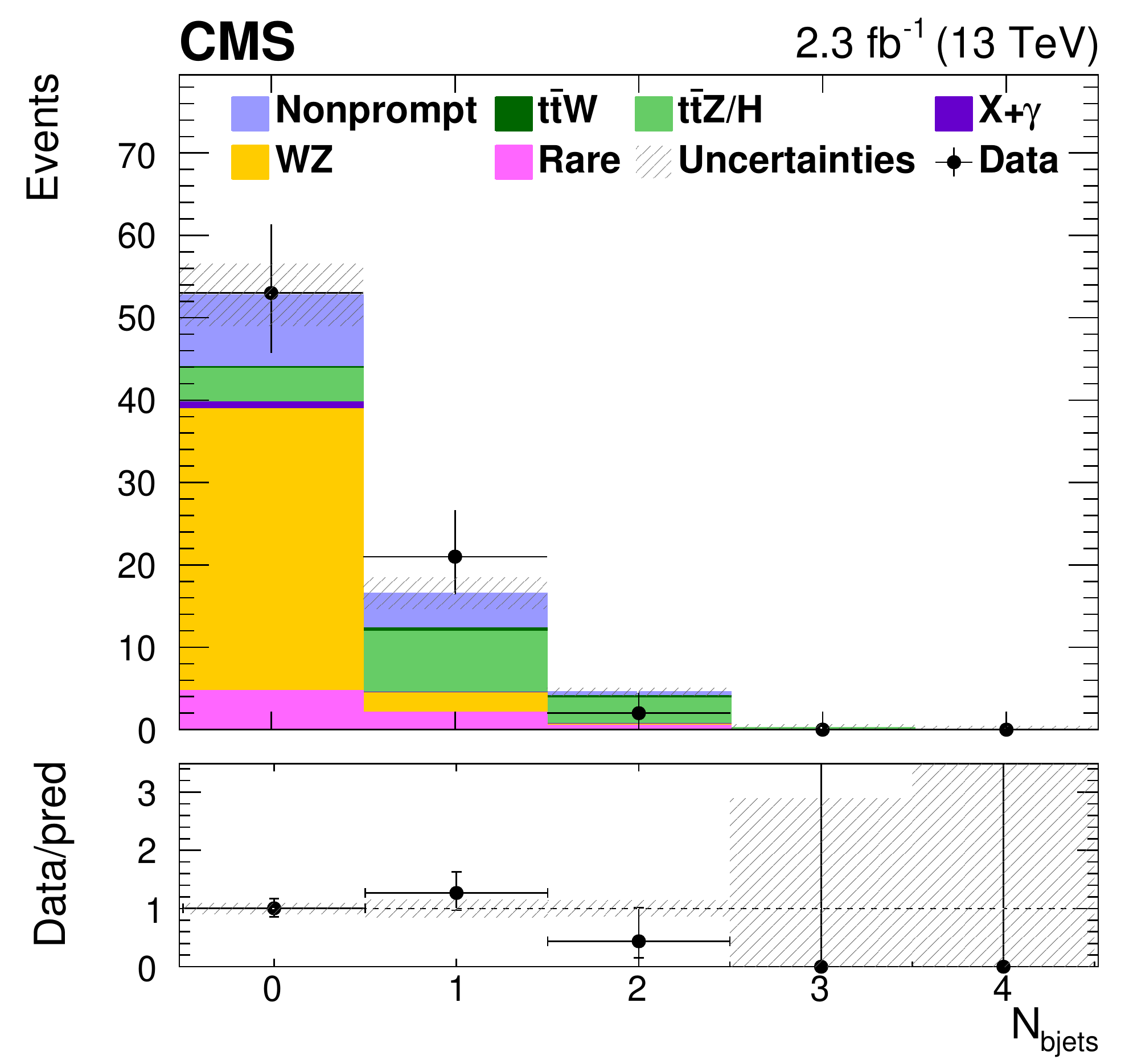}\\
	\includegraphics[width=.33\textwidth]{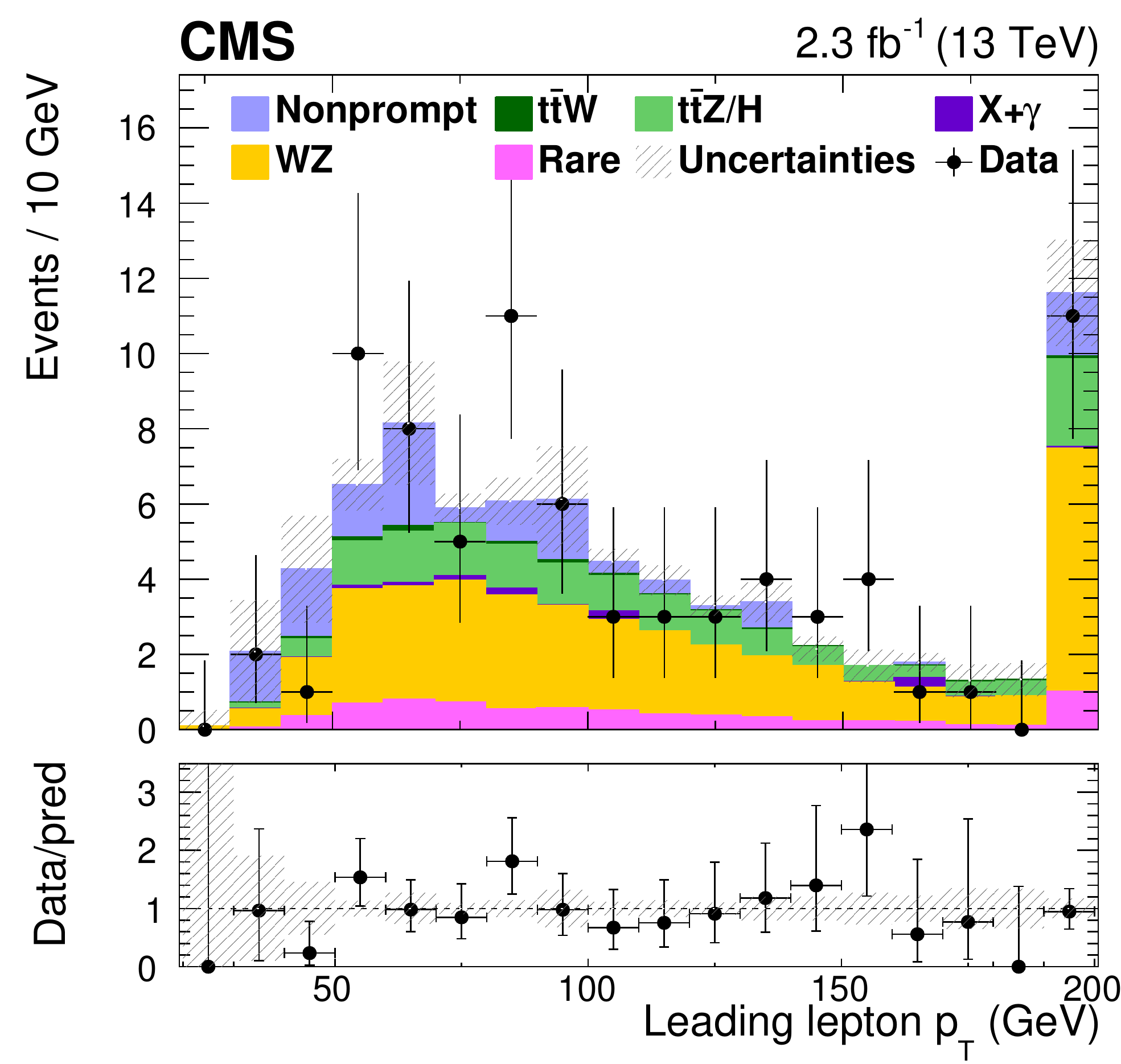}
	\includegraphics[width=.33\textwidth]{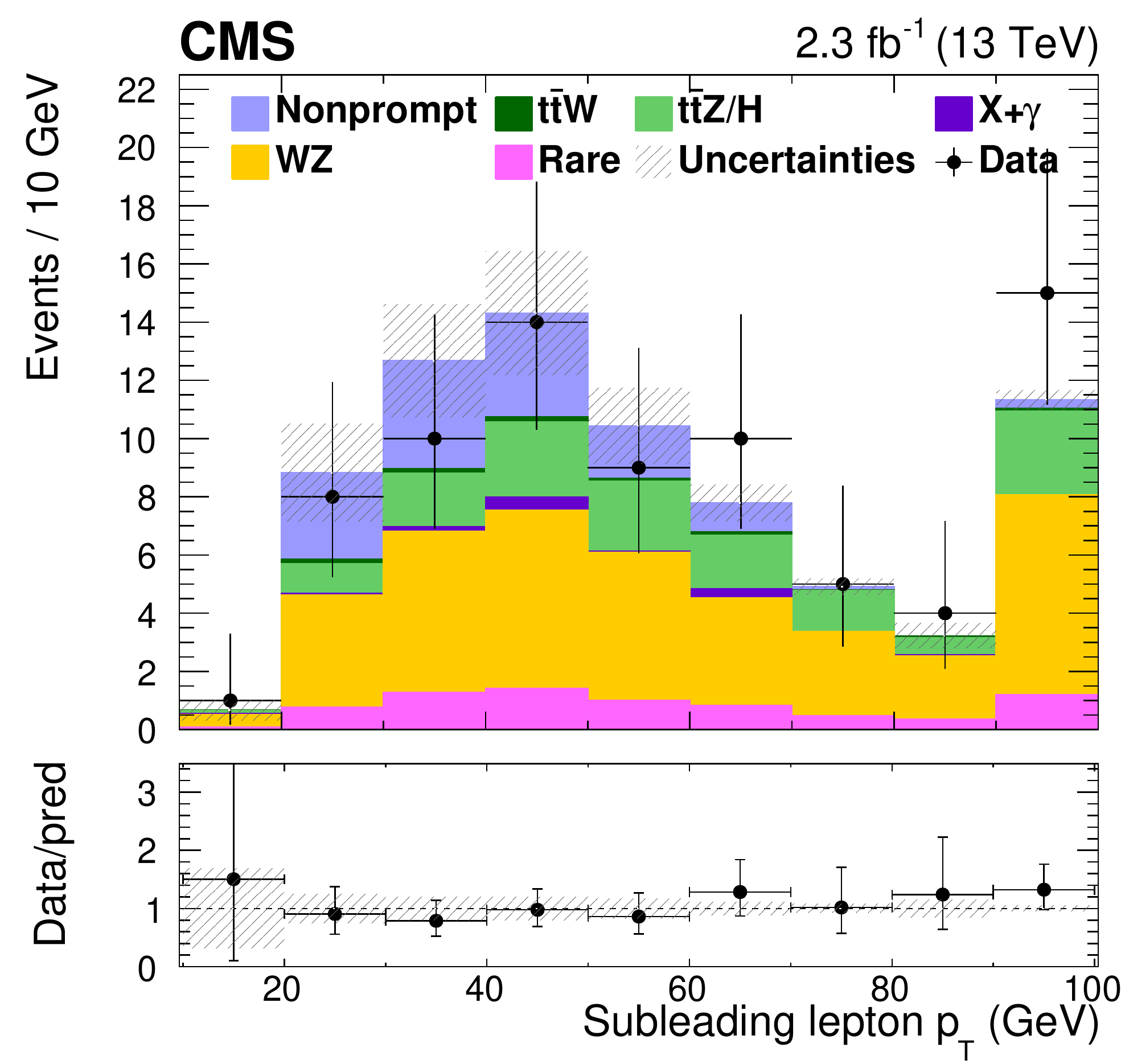}\\
	\includegraphics[width=.33\textwidth]{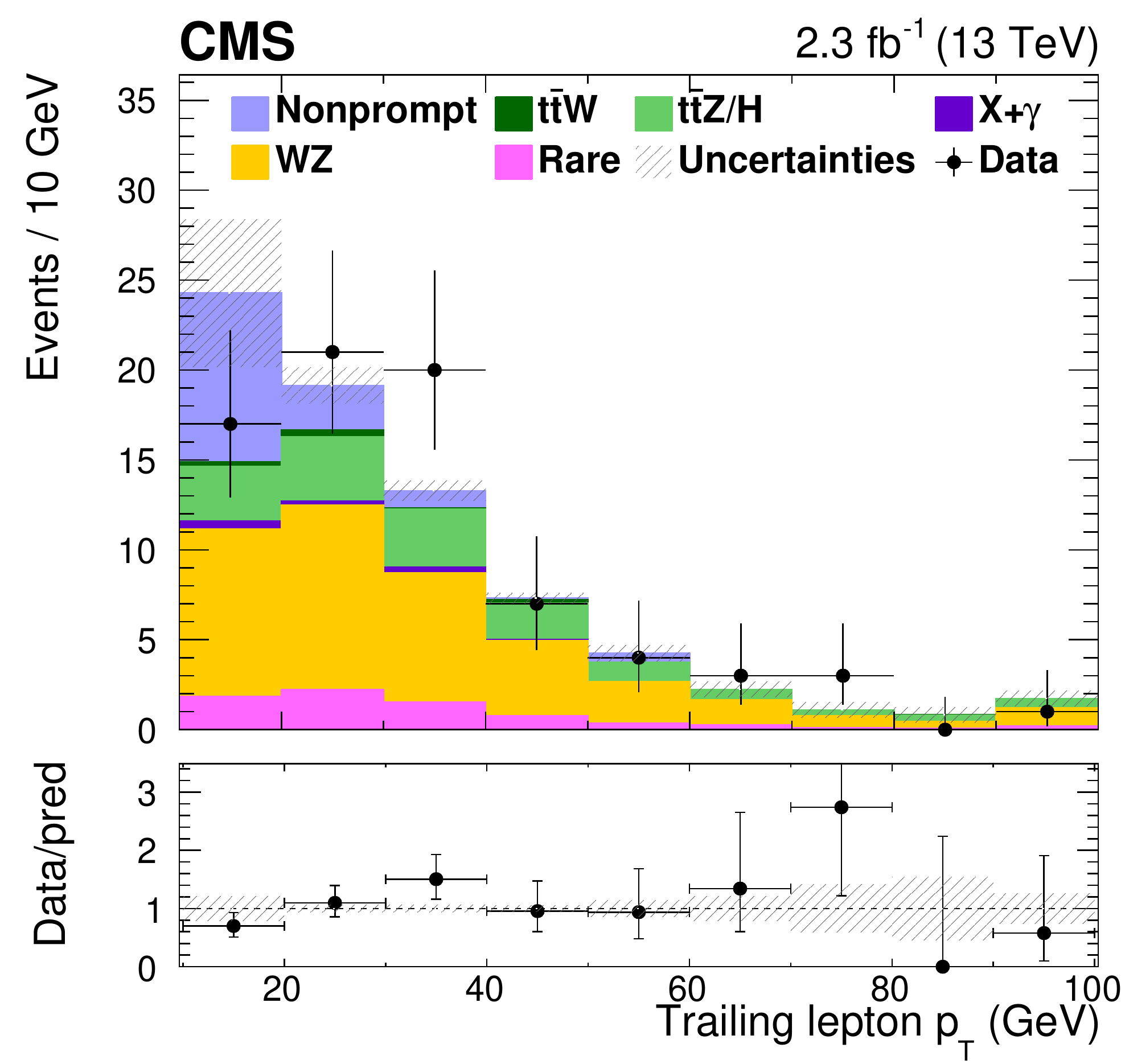}
	\includegraphics[width=.33\textwidth]{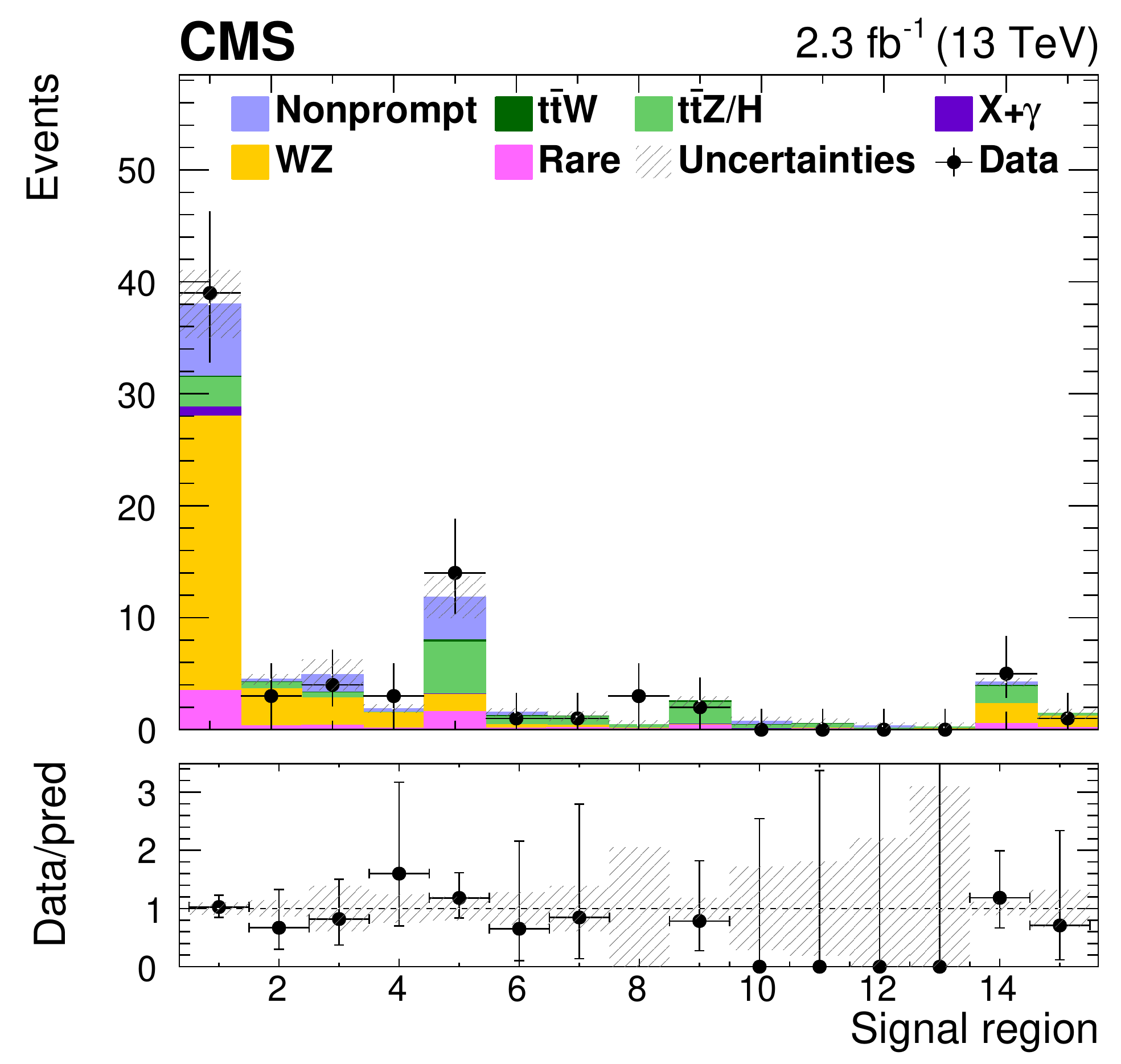}
\caption{On-Z samples: from left to right, top to bottom,  distributions of  $\HT$,  $\ptmiss$,  $\Njets$,  $\Nbjets$, $\pt$ of leptons for the predicted backgrounds and for the data in the on-Z baseline selection region, in these plots the rightmost bin contains the overflow from counts outside the range of the plot. On the bottom-right corner the total predicted background and the number of events observed in the 15 on-Z SRs is shown.}
\label{fig:fullOnZ}
\end{figure*}

Expected event yields are compared to the observation in Tables~\ref{tab:offZfull2} and~\ref{tab:onZfull2}. Comparisons of distributions of $\HT$, $\ptmiss$, $\Njets$, $\Nbjets$, leading lepton $\pt$, subleading lepton $\pt$, and trailing lepton $\pt$ measured in data with those predicted by the background estimation methods are shown in Fig.~\ref{fig:fullOffZ} (Fig.~\ref{fig:fullOnZ}), using all the events satisfying the off-Z (on-Z) SR selection criteria. The nonprompt lepton background comes from the technique described in Section~\ref{sec:backgrounds}. The hatched band represents the total background uncertainty in each bin. A graphical summary of predicted backgrounds and observed event yields in individual SRs is also shown. In these figures, the ``rare'' component is the sum over several SM processes, such as triboson production, associated Higgs production, $\ttbar\ttbar$, and other lower cross section processes.

\begin{table*}[tbp!]
\topcaption{Off-Z SRs: Comparison of observed event yields in data with predicted background yields.}
\label{tab:offZfull2}
\centering
\renewcommand{\arraystretch}{1.3}
\begin{tabular}{cccccc}\hline

\multirow{2}{*}{\Nbjets} 	& \multirow{2}{*}{$\HT$  (\GeVns{})}	& \multirow{2}{*}{$\ptmiss$ (\GeVns{})}	&  \multirow{2}{*}{Predicted}	&  \multirow{2}{*}{Observed}   &  \multirow{2}{*}{SR (off-Z)}          \\
							&								&			&	                       &  \\
\hline
\multirow{4}{*}{0 b-tags} 	& \multirow{2}{*}{60-400} 		& 50-150	&  $19.26^{+4.81}_{-4.80}$	& $18$                                                       & SR 1  \\ \cline{3-6}
							&								& 150-300	&  $1.16^{+0.31}_{-0.20}$	 & $4$       & SR 2   \\ \cline{2-6}
							& \multirow{2}{*}{400-600} 		& 50-150	&  $1.20^{+0.47}_{-0.40}$	& $3$	                             & SR 3   \\ \cline{3-6}
							& 								& 150-300	&  $0.29^{+0.44}_{-0.09}$	& $0$	     & SR 4   \\ \hline
\multirow{4}{*}{1 b-tags}   &   \multirow{2}{*}{60-400}		& 50-150	&  $16.57\pm4.52$			& $24$		                                             & SR 5   \\ \cline{3-6}
							&								& 150-300	&  $2.32^{+0.80}_{-0.76}$       & $1$	     & SR 6   \\ \cline{2-6}
   							&   \multirow{2}{*}{400-600}	& 50-150	&  $0.67^{+0.45}_{-0.09}$		& $2$ 	                             & SR 7   \\ \cline{3-6}
                            & 								& 150-300	&  $0.48^{+0.29}_{-0.07}$		& $0$ 	                             & SR 8   \\\hline
\multirow{4}{*}{2 b-tags}	&      \multirow{2}{*}{60-400} 	& 50-150	&  $4.49^{+1.81}_{-1.79}$		& $4$                                                        & SR 9   \\ \cline{3-6}
					&						& 150-300		&  $0.31^{+0.44}_{-0.09}$		& $1$ 			     & SR 10   \\ \cline{2-6}
					&  \multirow{2}{*}{400--600}	& 50--150			&  $0.40^{+0.27}_{-0.26}$		& $0$                                & SR 11   \\ \cline{3-6}
					&						& 150--300		&  $0.08^{+0.43}_{-0.08}$		& $0$ 		             & SR 12   \\ \hline
$ \geq$3 b-tags				&  60--600			& 50-300			&  $0.13^{+0.43}_{-0.09}$		& $0$ 		                     & SR 13   \\ \hline
$\geq$0 b-tags				&  $>$600				& 50-300 			&  $1.84^{+0.44}_{-0.37}$		& $3$ 	                     & SR 14   \\ \hline
$\geq$0 b-tags			&  $\geq$0				& $ \geq$300		&  $1.62^{+1.22}_{-1.19}$		& $0$ 		                             & SR 15   \\ \hline
\end{tabular}
\end{table*}

\begin{table*}[tbp!]
\topcaption{On-Z SRs: Comparison of observed event yields in data with predicted background yields.}
\label{tab:onZfull2}
\centering

\renewcommand{\arraystretch}{1.3}
\begin{tabular}{cccccc}\hline

\multirow{2}{*}{\Nbjets} 	& \multirow{2}{*}{$\HT$  (\GeVns{})}	& \multirow{2}{*}{$\ptmiss$ (\GeVns{})}	&  \multirow{2}{*}{Predicted}	&  \multirow{2}{*}{Observed}  &   \multirow{2}{*}{SR (on-Z)}          \\
							&										&										&		  						&								&    \\
\hline
\multirow{4}{*}{0 b-tags} 	& \multirow{2}{*}{60--400} 	& 50--150			&  $38.01\pm5.92$ 			& $39$		 & SR 1  \\ \cline{3-6}
					&                               & 150--300			&  $4.48^{+0.84}_{-0.75}$	& $3$		 & SR 2   \\ \cline{2-6}
					& \multirow{2}{*}{400--600} 	& 50--150			&  $4.88^{+1.49}_{-1.47}$	& $4$	         & SR 3   \\ \cline{3-6}
					& 				& 150--300			&  $1.88^{+0.47}_{-0.39}$	& $3$		 & SR 4   \\ \hline
\multirow{4}{*}{1 b-tags}   &   \multirow{2}{*}{60--400}	& 50--150			&  $11.84^{+2.28}_{-2.26}$	& $14$	                 & SR 5   \\ \cline{3-6}
                                 		&				& 150--300			&  $1.53^{+0.42}_{-0.34}$	& $1$	 & SR 6   \\ \cline{2-6}
   					&   \multirow{2}{*}{400--600}		& 50--150			&  $1.18^{+0.49}_{-0.23}$	& $1$ 	 & SR 7   \\ \cline{3-6}
                                 		& 				& 150--300			&  $0.42^{+0.44}_{-0.10}$	& $3$ 	 & SR 8   \\ \hline
\multirow{4}{*}{2 b-tags}	&  \multirow{2}{*}{60--400} 	& 50--150			&  $2.55^{+0.67}_{-0.51}$	& $2$                    & SR 9   \\ \cline{3-6}
					&				& 150--300			&  $0.72^{+0.76}_{-0.28}$	& $0$ 		 & SR 10   \\ \cline{2-6}
					&  \multirow{2}{*}{400--600}		& 50--150			&  $0.55^{+0.45}_{-0.13}$	& $0$ 	 & SR 11   \\ \cline{3-6}
					&				& 150--300			&  $0.31^{+0.51}_{-0.17}$	& $0$ 		 & SR 12   \\ \hline
$ \geq$3 b-tags				&  	60--600		& 50--300			&  $0.21^{+0.44}_{-0.13}$	& $0$ 			 & SR 13   \\ \hline
$\geq$0 b-tags				&  $>$600				& 50--300 			&  $4.22^{+0.68}_{-0.63}$	& $5$ 	 & SR 14   \\ \hline
$\geq$0 b-tags			&  $\geq$0					& $ \geq$300		&  $1.41^{+0.50}_{-0.25}$	& $1$ 		 & SR 15   \\ \hline
\end{tabular}
\end{table*}


The number of events observed in data is found to be consistent with predicted SM background yields. The results are used to calculate upper limits on the production cross section of gluinos or squarks for the various models discussed in Section~\ref{sec:strategy}, as a function of the gluino or squark, and the chargino or neutralino masses. For each mass hypothesis, the observation, background predictions, and expected signal yields from all on-Z and off-Z SRs are combined to extract an upper limit on the cross section, at 95\% confidence level (CL) using the asymptotic formulation of the LHC-style CL$_{\textrm s}$ method~\cite{Junk:1999kv,Read:2002hq,ATL-PHYS-PUB-2011-011,Cowan:2010js}. Log-normal nuisance parameters are used to describe the systematic uncertainties listed in Section~\ref{sec:systematics}.

\begin{figure*}[htbp]
\centering
	\includegraphics[width=.46\textwidth]{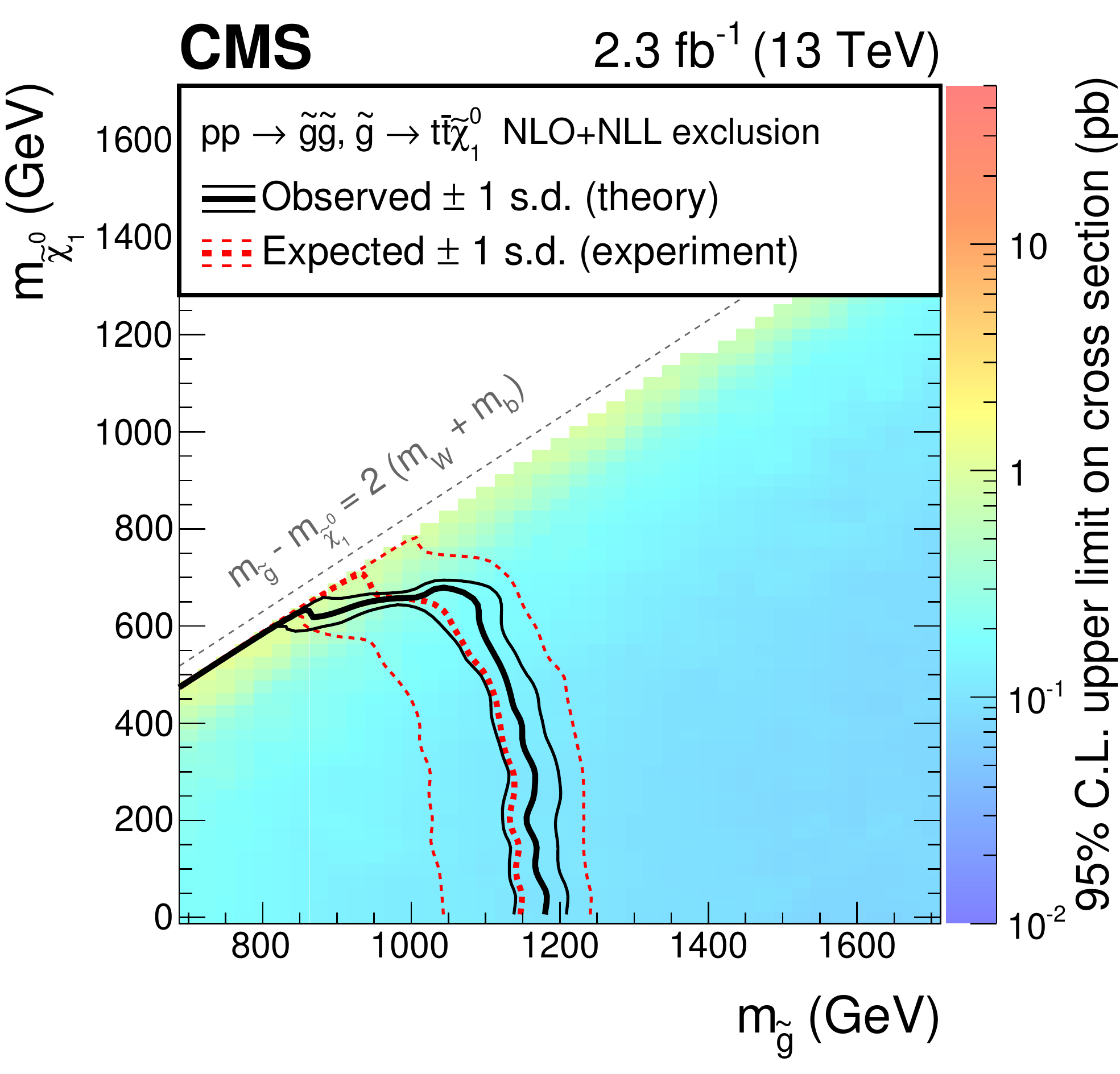}
	\includegraphics[width=.46\textwidth]{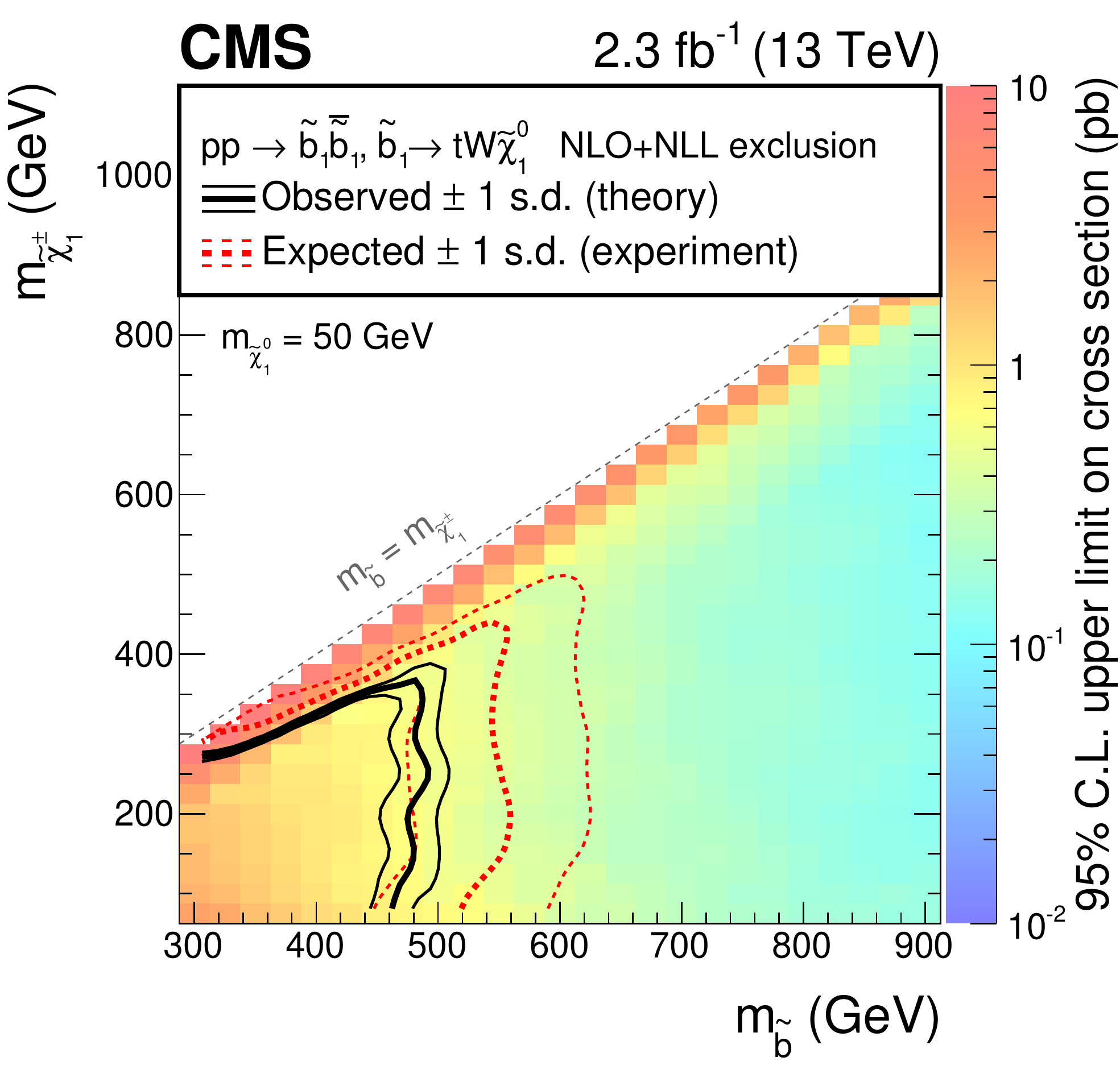}
	\includegraphics[width=.46\textwidth]{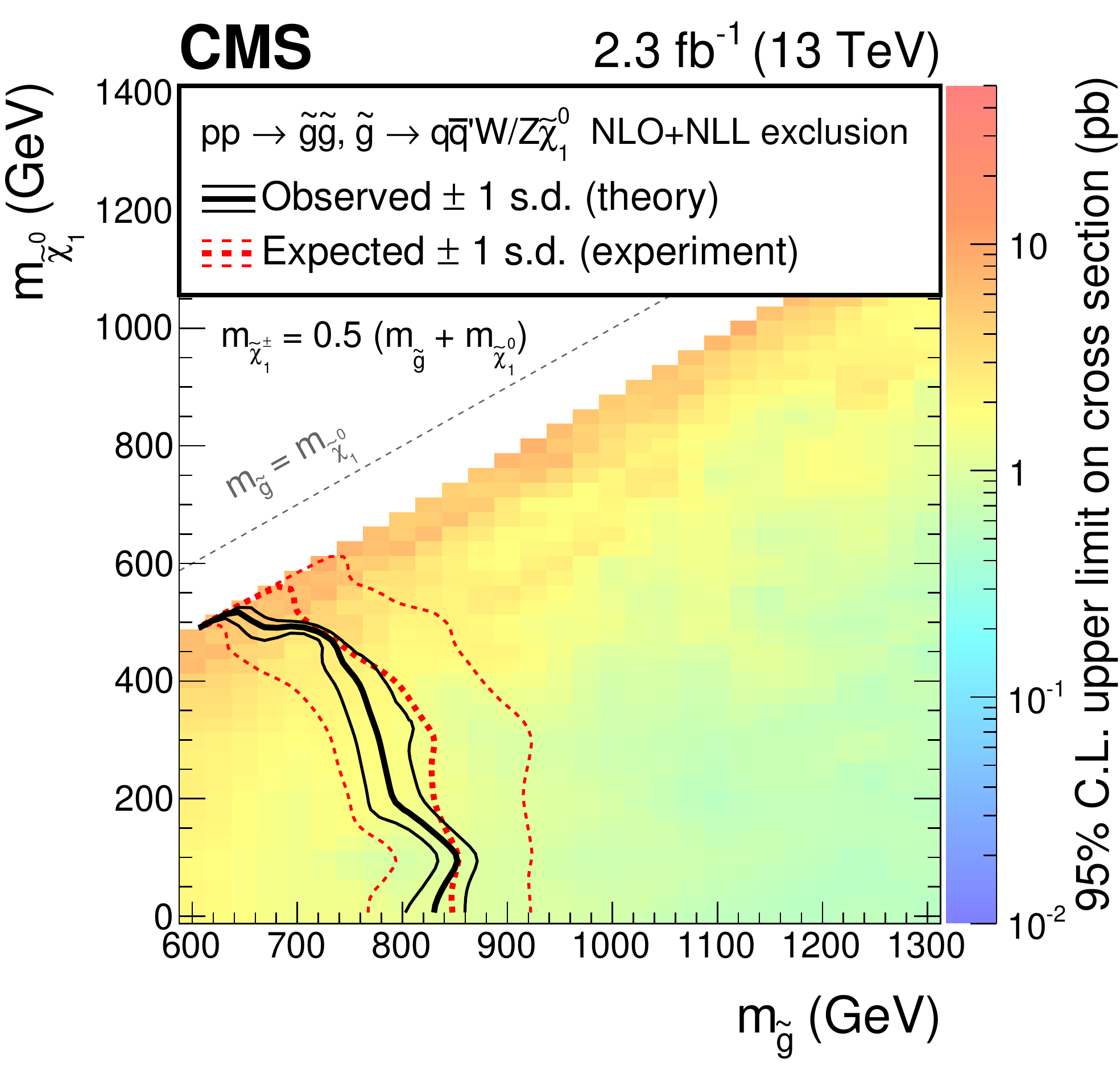}
\caption{\label{fig:exclusions}Exclusion contours as a function of $m_{\PSg}$ or $m_{\PSQb}$, and $m_{\chiz}$ or $m_{\PSGcpm}$, for the simplified SUSY models (top-left) T1tttt, (top-right) T6ttWW, and (bottom) T5qqqqWZ. The color scale indicates the 95\% CL observed upper limits on the cross section. The observed (expected) exclusion curves are indicated by the solid (dashed) lines using NLO+NLL production cross sections, along with the corresponding  $\pm$1 s.d. theoretical (experimental) uncertainties.}
\end{figure*}

{\tolerance=1200
These upper limits are used to calculate exclusion contours on the concerned sparticles mass plane, shown in Fig.~\ref{fig:exclusions} for the simplified models under consideration. In these figures, the thick black lines delineate the observed exclusion region, which is at the lower masses side. The uncertainty in the observed limit, represented by the thinner black lines, is the propagation of the NLO+NLL cross section uncertainties for the relevant signal process~\cite{Kulesza:2008jb,Kulesza:2009kq,Beenakker:2009ha,Beenakker:2011fu}. The red dashed lines represent the expected limits with the uncertainties reflecting those discussed in Section~\ref{sec:systematics}.
\par}

\begin{figure*}[htbp]
\centering
	\includegraphics[width=.46\textwidth]{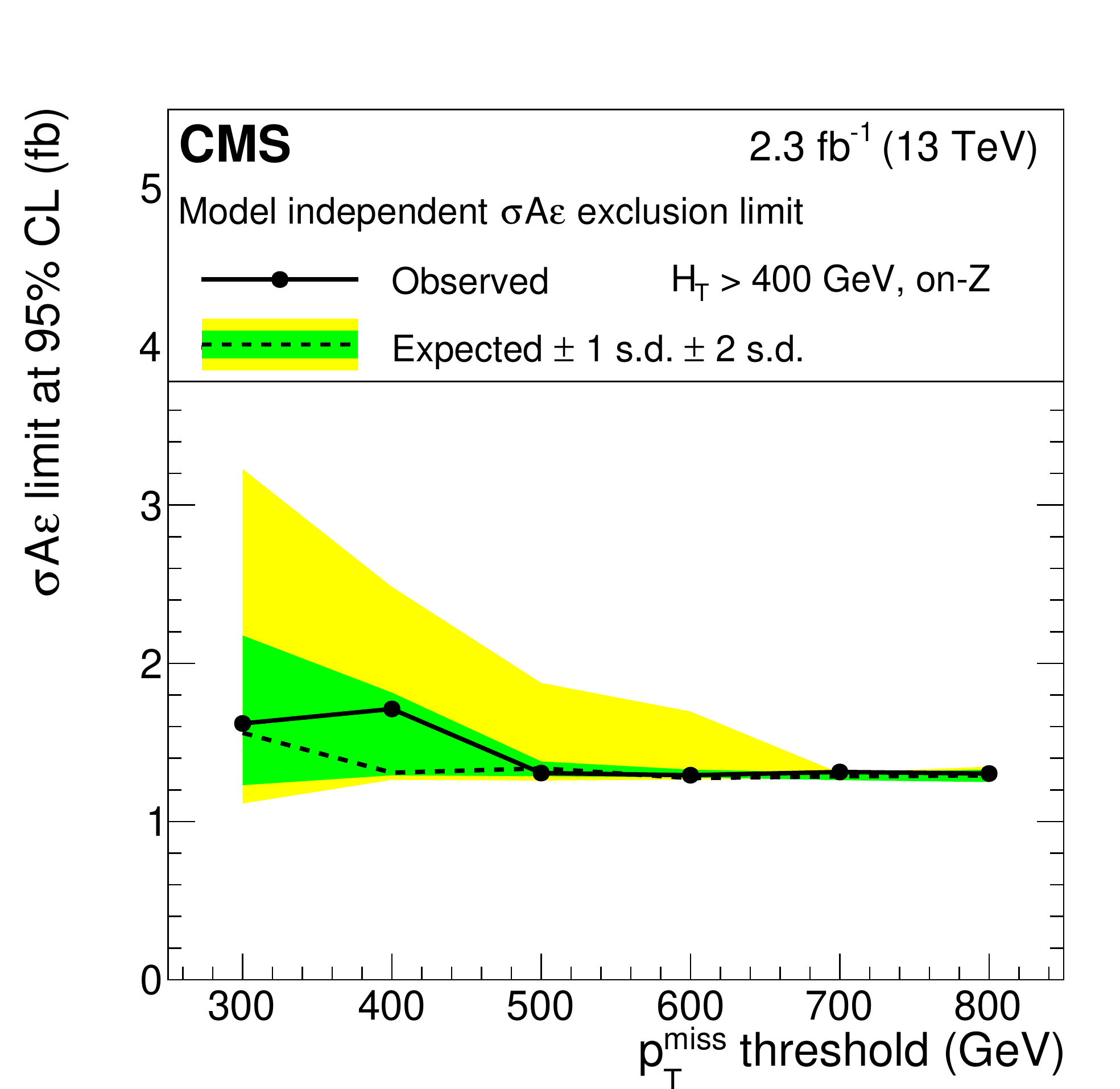}
	\includegraphics[width=.46\textwidth]{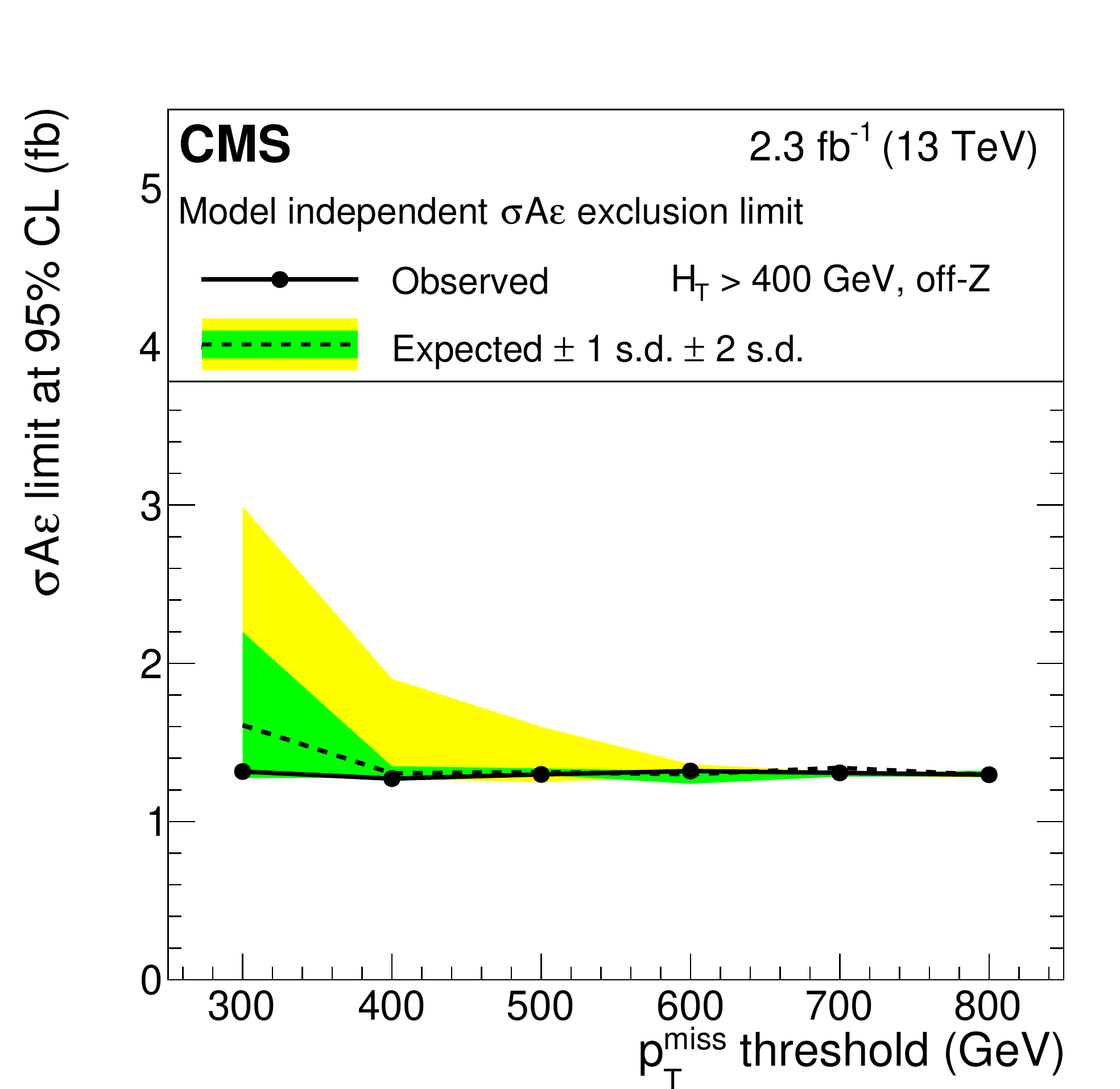}
\caption{\label{fig:modelindep} Limits on the product of cross section, detector acceptance, and selection efficiency, $\sigma A \epsilon$, for the production of multilepton events with (left) or without (right) an on-Z lepton pair as a function of the $\ptmiss$ threshold.}
\end{figure*}

The yields and background predictions can be used to test additional BSM physics scenarios. To facilitate such reinterpretations, we provide limits on the number of multilepton events as a function of the $\ptmiss$ threshold in the kinematic tails of this search. These limits are obtained based on the tails of our SRs, in particular we consider events with $\HT > 400\GeV$, both with and without an on-Z lepton pair, employing the LHC-style CL$_{\textrm s}$ method carried out with pseudo-experiments~\cite{Junk:1999kv,Read:2002hq,ATL-PHYS-PUB-2011-011}. They are shown in Fig.~\ref{fig:modelindep} in terms of the product of cross section ($\sigma$), detector acceptance ($A$), and selection efficiency ($\epsilon$). As we increase the $\ptmiss$ threshold, the observed and expected limits converge to 1.3\unit{fb}.

\section{Summary}
\label{sec:conclusions}
We have presented the search for beyond-the-standard-model physics in final states with at least 3 leptons, electrons or muons, using proton-proton data collected with the CMS detector at $\sqrt{s} = 13\TeV$, corresponding to an integrated luminosity of \mllumi. The analysis makes use of techniques based on control samples in data to estimate reducible backgrounds and to validate the simulation for use in estimating irreducible backgrounds. To maximize sensitivity to a broad range of possible signal models, we investigate 30 exclusive signal regions. The event yields observed in data are in agreement with the standard model background predictions.

This search is designed to be sensitive to multiple BSM models. As an example, we interpret the result in the context of a gluino-pair production model that features cascade decays producing four top quarks in the final state. In this simplified model, we exclude gluinos with a mass of up to 1175\GeV in the case of a massless lightest supersymmteric particle (LSP). For gluino masses up to approximately 1150\GeV, neutralino masses below 650\GeV are excluded. These are the first CMS results reported in this final state at $\sqrt{s} = 13\TeV$.

In a bottom squark pair production model with cascade decays that contain two top quarks and two additional \Wpm bosons, we also set limits on the masses of the bottom squark and the chargino. We exclude bottom squarks with a mass of up to 450\GeV in the case of a chargino with a mass of 200\GeV. For bottom squark masses up to approximately 450\GeV, neutralino masses below 300\GeV are excluded. In a similar search at $\sqrt{s} = 8\TeV$~\cite{Khachatryan:1976453}, the bottom squark mass limit was slightly larger and the chargino mass limit was approximately the same.

An additional interpretation is presented in a gluino pair production model with four light quarks and two vector bosons in the final state. For the case of one W and one Z boson in the final state, we exclude gluino masses up to 825\GeV when the LSP mass is 100\GeV, and LSP masses up to 500\GeV for 700\GeV gluinos.

Finally, limits on the number of multilepton events with $\HT > 400$\GeV as a function of $\ptmiss$ threshold are also presented in terms of the product of cross section, detector acceptance, and selection efficiency. For a $\ptmiss$ threshold greater than 500\GeV, the observed and expected limits are 1.3\unit{fb}.

\begin{acknowledgments}
\hyphenation{Bundes-ministerium Forschungs-gemeinschaft Forschungs-zentren} We congratulate our colleagues in the CERN accelerator departments for the excellent performance of the LHC and thank the technical and administrative staffs at CERN and at other CMS institutes for their contributions to the success of the CMS effort. In addition, we gratefully acknowledge the computing centres and personnel of the Worldwide LHC Computing Grid for delivering so effectively the computing infrastructure essential to our analyses. Finally, we acknowledge the enduring support for the construction and operation of the LHC and the CMS detector provided by the following funding agencies: the Austrian Federal Ministry of Science, Research and Economy and the Austrian Science Fund; the Belgian Fonds de la Recherche Scientifique, and Fonds voor Wetenschappelijk Onderzoek; the Brazilian Funding Agencies (CNPq, CAPES, FAPERJ, and FAPESP); the Bulgarian Ministry of Education and Science; CERN; the Chinese Academy of Sciences, Ministry of Science and Technology, and National Natural Science Foundation of China; the Colombian Funding Agency (COLCIENCIAS); the Croatian Ministry of Science, Education and Sport, and the Croatian Science Foundation; the Research Promotion Foundation, Cyprus; the Secretariat for Higher Education, Science, Technology and Innovation, Ecuador; the Ministry of Education and Research, Estonian Research Council via IUT23-4 and IUT23-6 and European Regional Development Fund, Estonia; the Academy of Finland, Finnish Ministry of Education and Culture, and Helsinki Institute of Physics; the Institut National de Physique Nucl\'eaire et de Physique des Particules~/~CNRS, and Commissariat \`a l'\'Energie Atomique et aux \'Energies Alternatives~/~CEA, France; the Bundesministerium f\"ur Bildung und Forschung, Deutsche Forschungsgemeinschaft, and Helmholtz-Gemeinschaft Deutscher Forschungszentren, Germany; the General Secretariat for Research and Technology, Greece; the National Scientific Research Foundation, and National Innovation Office, Hungary; the Department of Atomic Energy and the Department of Science and Technology, India; the Institute for Studies in Theoretical Physics and Mathematics, Iran; the Science Foundation, Ireland; the Istituto Nazionale di Fisica Nucleare, Italy; the Ministry of Science, ICT and Future Planning, and National Research Foundation (NRF), Republic of Korea; the Lithuanian Academy of Sciences; the Ministry of Education, and University of Malaya (Malaysia); the Mexican Funding Agencies (BUAP, CINVESTAV, CONACYT, LNS, SEP, and UASLP-FAI); the Ministry of Business, Innovation and Employment, New Zealand; the Pakistan Atomic Energy Commission; the Ministry of Science and Higher Education and the National Science Centre, Poland; the Funda\c{c}\~ao para a Ci\^encia e a Tecnologia, Portugal; JINR, Dubna; the Ministry of Education and Science of the Russian Federation, the Federal Agency of Atomic Energy of the Russian Federation, Russian Academy of Sciences, and the Russian Foundation for Basic Research; the Ministry of Education, Science and Technological Development of Serbia; the Secretar\'{\i}a de Estado de Investigaci\'on, Desarrollo e Innovaci\'on and Programa Consolider-Ingenio 2010, Spain; the Swiss Funding Agencies (ETH Board, ETH Zurich, PSI, SNF, UniZH, Canton Zurich, and SER); the Ministry of Science and Technology, Taipei; the Thailand Center of Excellence in Physics, the Institute for the Promotion of Teaching Science and Technology of Thailand, Special Task Force for Activating Research and the National Science and Technology Development Agency of Thailand; the Scientific and Technical Research Council of Turkey, and Turkish Atomic Energy Authority; the National Academy of Sciences of Ukraine, and State Fund for Fundamental Researches, Ukraine; the Science and Technology Facilities Council, UK; the US Department of Energy, and the US National Science Foundation.

Individuals have received support from the Marie-Curie programme and the European Research Council and EPLANET (European Union); the Leventis Foundation; the A. P. Sloan Foundation; the Alexander von Humboldt Foundation; the Belgian Federal Science Policy Office; the Fonds pour la Formation \`a la Recherche dans l'Industrie et dans l'Agriculture (FRIA-Belgium); the Agentschap voor Innovatie door Wetenschap en Technologie (IWT-Belgium); the Ministry of Education, Youth and Sports (MEYS) of the Czech Republic; the Council of Science and Industrial Research, India; the HOMING PLUS programme of the Foundation for Polish Science, cofinanced from European Union, Regional Development Fund, the Mobility Plus programme of the Ministry of Science and Higher Education, the National Science Center (Poland), contracts Harmonia 2014/14/M/ST2/00428, Opus 2013/11/B/ST2/04202, 2014/13/B/ST2/02543 and 2014/15/B/ST2/03998, Sonata-bis 2012/07/E/ST2/01406; the Thalis and Aristeia programmes cofinanced by EU-ESF and the Greek NSRF; the National Priorities Research Program by Qatar National Research Fund; the Programa Clar\'in-COFUND del Principado de Asturias; the Rachadapisek Sompot Fund for Postdoctoral Fellowship, Chulalongkorn University and the Chulalongkorn Academic into Its 2nd Century Project Advancement Project (Thailand); and the Welch Foundation, contract C-1845.
\end{acknowledgments}
\bibliography{auto_generated}

\cleardoublepage \appendix\section{The CMS Collaboration \label{app:collab}}\begin{sloppypar}\hyphenpenalty=5000\widowpenalty=500\clubpenalty=5000\textbf{Yerevan Physics Institute,  Yerevan,  Armenia}\\*[0pt]
V.~Khachatryan, A.M.~Sirunyan, A.~Tumasyan
\vskip\cmsinstskip
\textbf{Institut f\"{u}r Hochenergiephysik,  Wien,  Austria}\\*[0pt]
W.~Adam, E.~Asilar, T.~Bergauer, J.~Brandstetter, E.~Brondolin, M.~Dragicevic, J.~Er\"{o}, M.~Flechl, M.~Friedl, R.~Fr\"{u}hwirth\cmsAuthorMark{1}, V.M.~Ghete, C.~Hartl, N.~H\"{o}rmann, J.~Hrubec, M.~Jeitler\cmsAuthorMark{1}, A.~K\"{o}nig, I.~Kr\"{a}tschmer, D.~Liko, T.~Matsushita, I.~Mikulec, D.~Rabady, N.~Rad, B.~Rahbaran, H.~Rohringer, J.~Schieck\cmsAuthorMark{1}, J.~Strauss, W.~Waltenberger, C.-E.~Wulz\cmsAuthorMark{1}
% ...
\vskip\cmsinstskip
% ...
1:~~Also at Vienna University of Technology, Vienna, Austria\\
% ...

\end{sloppypar}
\end{document}